\documentclass[12pt]{iopart}
\usepackage{graphicx}
\usepackage{setspace}
\usepackage{iopams}  
\usepackage{tikz}
\expandafter\let\csname equation*\endcsname\relax
\expandafter\let\csname endequation*\endcsname\relax
\usepackage{amsmath} 
\usepackage[labelfont={small},subrefformat=parens,caption=false]{subfig}
\usepackage{amssymb,amsbsy,amsfonts,bm} %
\usepackage{gensymb}
\usepackage{bbm}
\usepackage{url} 
\usepackage{nicefrac}
\usepackage{xcolor}
\usepackage{xspace}
\usepackage{soul} 
\usepackage{cite} 
\usepackage[
        pdfencoding=auto,%
        colorlinks=true,%
        linkcolor=blue,%
        citecolor=blue, %
        urlcolor=blue,
				breaklinks=true]{hyperref}
\usepackage[hyphenbreaks=true]{breakurl}
\usepackage{array}

\usepackage{doi} 
\renewcommand{\vec}{\boldsymbol}

\DeclareMathOperator*{\argmax}{arg\,max}

\newcommand{\pr}{p}
\newcommand{\given}{\,|\,}  

\makeatletter
\newcommand\newsubcommand[3]{\newcommand#1{#2\sc@sub{#3}}}
\def\sc@sub#1{\def\sc@thesub{#1}\@ifnextchar_{\sc@mergesubs}{_{\sc@thesub}}}
\def\sc@mergesubs_#1{_{\sc@thesub#1}}

\newcommand\newsupcommand[3]{\newcommand#1{#2\sc@sup{#3}}}
\def\sc@sup#1{\def\sc@thesup{#1}\@ifnextchar^{\sc@mergesups}{^{\sc@thesup}}}
\def\sc@mergesups^#1{^{\sc@thesup#1}}
\makeatother

\newcommand{\inputvec}{\mathbf}
\newcommand{\ordervec}{\vec}

\newcommand{\genobs}{y}

\newsubcommand{\genobsvec}{\ordervec{\genobs}}{k}
\newsubcommand{\genobsvecset}{\ordervec{\inputvec{\genobs}}}{k}
\newcommand{\genobsset}{\inputvec{y}}
\newcommand{\genobsth}{\genobs_{\textup{th}}}          
\newcommand{\design}{\mathbf{x}}

\newcommand{\abar}{\sigma_a} 

\newcommand{\normal}{\mathcal{N}}
\newcommand{\inputdimvec}{}
\newcommand{\kinparvec}{\inputdimvec{x}}
\newcommand{\qoi}{\mathbf{Q}}
\newcommand{\observations}{\mathbf{D}}
\newcommand{\shrinkage}{\mathcal{S}}
\newcommand{\omegalab}{\omega_{\text{lab}}}
\newcommand{\thetalab}{\theta_{\text{lab}}}
\newcommand{\dsigma}{\dd{\sigma}}
\newcommand{\diffcs}{\dsigma}  

\newcommand{\eg}{\textit{e.g.}\xspace}

\newcommand{\dd}{\mathrm{d}}

\begin{document}

\title[Bayesian Framework for Model Uncertainties]{Get on the BAND Wagon: A Bayesian Framework for Quantifying Model Uncertainties in Nuclear Dynamics}

\author{D.~R.~Phillips$^{1}$, R.~J.\ Furnstahl$^{2}$, U.\ Heinz$^{2}$, T.\ Maiti$^{3}$, 
W.~Nazarewicz$^{4}$, F.~M. Nunes$^{4}$, M.\ Plumlee$^{5,6}$, M.~T.~Pratola$^{7}$, S.\ Pratt$^4$, F.~G.\ Viens$^{3}$, S.~M.\ Wild$^{6,8}$}
\address{$^{1}$Department of Physics and Astronomy and Institute of Nuclear and Particle Physics, Ohio University, Athens, OH 45701, USA \\
$^{2}$Department of Physics, The Ohio State University, Columbus, OH 43210, USA \\
$^{3}$Department of Statistics and Probability, Michigan State University, East Lansing, Michigan 48824, USA \\
$^{4}$Department of Physics and Astronomy and Facility for Rare Isotope Beams, Michigan State University, East Lansing, Michigan 48824, USA \\
$^5$Department of Industrial Engineering and Management Sciences, Northwestern University, Evanston, Illinois 60208, USA \\
$^6$NAISE, Northwestern University, Evanston, Illinois 60208, USA \\
$^{7}$Department of Statistics, The Ohio State University, Columbus, OH 43210, USA \\
$^8$Mathematics and Computer Science Division, Argonne National Laboratory, Lemont, Illinois 60439, USA \\
}

\ead{phillid1@ohio.edu}

\vspace{10pt}
\begin{indented}
\item[]\today 
\end{indented} 

\begin{abstract}
We describe the Bayesian Analysis of Nuclear Dynamics (BAND) framework, a cyberinfrastructure that we are developing which will unify the treatment of nuclear models, experimental data, and associated uncertainties. We overview the statistical principles and nuclear-physics contexts underlying the BAND toolset, with an emphasis on Bayesian methodology's ability to leverage insight from multiple models. In order to facilitate understanding of these tools we provide a simple and accessible example of the BAND framework's application. Four case studies are presented to highlight how elements of the framework will enable progress on complex, far-ranging problems in nuclear physics. By collecting notation and terminology, providing illustrative examples, and giving an overview of the associated techniques, this paper aims to open paths through which the nuclear physics and statistics communities can contribute to and build upon the BAND framework.
\end{abstract}

\newpage
\tableofcontents

\markboth{The BAND Framework}{The BAND Framework}

%
%
%
%
%

\section{Introduction}\label{Intro}

\noindent
Progress in the theory of nuclei and nuclear matter has produced a multitude of models that describe extant data well. The atomic nucleus is a complex system and these models---many of which involve advanced numerical  simulation---provide essential insights into many nuclear-physics phenomena. The need for validation, verification, and uncertainty quantification of models that simulate real-world physical processes is a theme that is common to all physical sciences. As eloquently stated in the recent report \cite{NAS-Models} ``regardless of their underlying mathematical formalism or their intended purpose, [the complex models] share a common feature---they are not reality." In order to understand and use the results of nuclear-physics simulations well we must follow best practices for  statistical modeling and uncertainty quantification \cite{Saltelli2020}. All this means we are at an inflection point in  how nuclear-physics data should be analyzed: predictions and quantified uncertainties must use the collective wisdom of the best models, constrained by data, and include a unified treatment of {\it all} uncertainties.

Bayesian Analysis of Nuclear Dynamics (BAND) will be a set of publicly-available software tools---a cyberinfrastructure framework---designed to facilitate principled uncertainty quantification (UQ) with multiple nuclear models. It will enable reliable predictions for experimentally inaccessible environments, such as the properties and dynamics of matter at the core of neutron stars or in the first microseconds after the Big Bang. And it will make possible quantitative evaluation of the impact of new experiments, thus facilitating optimal use of investment in this science. 


\begin{table}[p   ]
\caption{\label{tab:glossary}Lexicon: {\it When I use a word it means what I choose it to mean, neither more nor less}~\cite{Carroll1871}.
        Note that several terms that are defined in the text of the article are not listed here. Instead, this table focuses on terms at the nuclear-physics/statistics interface whose use may otherwise cause confusion.}
\begin{tabular}{>{\itshape}p{0.25\textwidth}p{0.70\textwidth}}
\br 
\multicolumn{1}{l}{Term} & Usage here \\
\mr 
Calibration dataset& The \textit{observables} that are used to constrain the \textit{model parameters} \\
Computational tool & A piece of software that accomplishes a statistical or other data analysis task for a \textit{physics model} or a set of \textit{physics models}\\
Dataset & A collection of \textit{observables}\\
Domain scientist & Here, the nuclear physicist \\
Emulator & A computationally inexpensive way to interpolate results of an expensive \textit{physics model}  in its many-dimensional parameter space\\
Experimental design & The process of selecting amongst experimental options based on the optimization of a selected utility function\\
Experiments & Measurements in the nuclear laboratory \\
Framework & A set of inter-linked \textit{input tools} and 
\textit{computational tools} that can be used separately, or in concert\\
Input tool & An interrogative process by which the elements of the statistical analysis being carried out are established\\
Model & The combination of a \textit{physics model}, a \textit{calibration dataset}, and a
\textit{statistical model}\\
Model results & The probability distribution function obtained for \textit{observables} in the \textit{model}\\
Hyper\-parameter &  Parameter describing a prior distribution (Bayesian statistics usage)\\
Model parameters & Variables internal to the \textit{model} [Their (joint) probability distribution can be estimated from Bayesian statistics or otherwise learned from experiment through repeated parameter estimation]\\
Observables & The results of measurements described by \textit{physics models} \\
Physics model & The physical description of the \textit{observables}
through mathematical equations encoding physical rules and principles [These equations involve parameters that are usually  constrained by the \textit{calibration dataset}]\\
Predictions & Values obtained in the \textit{model} for \textit{observables} that are not part of the training dataset\\
Statistical model & The statistical framework to assess 
deficiencies of the \textit{physics model} and the uncertainties inherent in its predictions \\
\br
\end{tabular}
\end{table}

Contemporary nuclear physics involves statistical inference within complex and computationally intensive theoretical models that combine heterogeneous datasets taken at experimental facilities around the world. Modern UQ can enhance the predictive power of these models and optimize knowledge extraction from new measurements and observations. The goal of BAND is to translate novel statistical methods of UQ into software tools that address prominent current problems in nuclear physics (NP). This, in turn, will inform  near- and medium-term planning for experimental programs at leading NP facilities. This  interweaving of  statistical approaches into the dialog between nuclear physicists and experimental data will accelerate the theory-experiment feedback loop \cite{Bal14,Nazarewicz_2016} and lead to sustained innovation.

BAND will do all this by providing to the community a suite of codes that produce emulators for forefront, computationally-intensive nuclear models, and perform principled UQ that calibrates those models against data. Codes already exist---some publicly available, some written by members of our team and as yet unpublished---that implement parts of this UQ methodology. But BAND will go further. Because it is built on
Bayesian statistical methodology, it will also include a software tool to mix different models, thereby providing a multi-model prediction~\footnote{%
    Here and below, the term prediction refers to an observable that is an output of the Bayesian model but is not part of the dataset used to constrain the model. Our predictions therefore include quantities that have already been measured (i.e., what are sometimes called postdictions).} 
for key observables. This will permit the use of Bayesian Model Mixing for the quantitative assessment of model-related uncertainties in the multi-model context. A model-mixed prediction that enriches the physics and provides a full assessment of the modeling uncertainty of predictions is a natural outcome \cite{tebaldi2007,smith2009bayesian} within BAND. That prediction includes experimental {\em and} modeling errors, thus providing a unified statistical treatment of all uncertainties. Model-mixed predictions can then give insight into what experimental information will best constrain models.

To illustrate the power of this approach we take the example of the Facility for Rare Isotope Beams (FRIB) \cite{aFRIB}, which will come online soon and provide a wealth of new data on atomic nuclei and their reactions. A key physics target for FRIB is a quantitative understanding of the astrophysical rapid neutron capture (r-)process by which  many heavy elements such as gold and uranium are formed. This requires knowledge of the masses, decays, and reaction rates of short-lived neutron-rich nuclei. While FRIB will be able to produce many key r-process isotopes, it cannot measure {\it all} of the ${\approx\,}$3,000 nuclei involved. Nuclear-structure models, informed by the existing experimental datasets augmented by the new FRIB data, will have to carry out massive  extrapolations to provide the needed input for nucleosynthesis simulations \cite{Horowitz2019}. 

The arrival of the era of multi-messenger astrophysics~\cite{NSF_BIG_IDEAS,Meszaros2019} presents both an opportunity and a challenge for FRIB's program. The extrapolations needed  to interpret the  different signals from an extreme stellar event (e.g., neutrinos, optical, X-ray and gamma spectra, gravitational waves) require proper propagation of not just measurement errors, but also theoretical uncertainties. It is important that multi-messenger astrophysics---and other fields that need data on unstable nuclei---achieve the most possible benefit from FRIB. Guidance will be needed to optimize FRIB's precious beam: we need to assess which measurements might best reduce extrapolation errors for the properties outside experimental reach that affect the multi-messenger signal---or some other application of interest. This guidance should coherently use the information from different nuclear models and must account for theoretical uncertainties. 
 
BAND will also advance the  modeling of neutron stars and supernovae by assimilating new experimental information on exotic nuclei from FRIB and from high-energy heavy-ion collisions at RHIC \cite{RHIC} and the LHC \cite{LHC}. There are many other examples of potential framework applications, including critically needed quantified predictions for  tonne-scale experiments searching for the neutrinoless double-beta decay of nuclei \cite{Engel2018} as a definitive sign of new physics.


This article introduces the BAND software framework for multiple models in physics. (Further details on the framework can be found at the project webpage~\cite{BAND}.) Here we lay out a strategy for the use of Bayesian methods to assess model uncertainty in the nuclear-physics context. In order to ground that strategy in a common language and practice we provide guidance on the use of Bayesian methods to the nuclear-physics community. The most novel sections of the paper are those pertaining to Bayesian Model Averaging (BMA) and the more general technique of Bayesian Model Mixing (BMM). While BMA is the most obvious (Bayesian) way to assess model uncertainty and is frequently employed, we strongly emphasize that it has important shortcomings which could be damaging in the nuclear-physics context. We therefore exhort nuclear physicists to focus on the more general BMM. We also present several nuclear-physics examples that illustrate the ways in which BAND could advance the field.

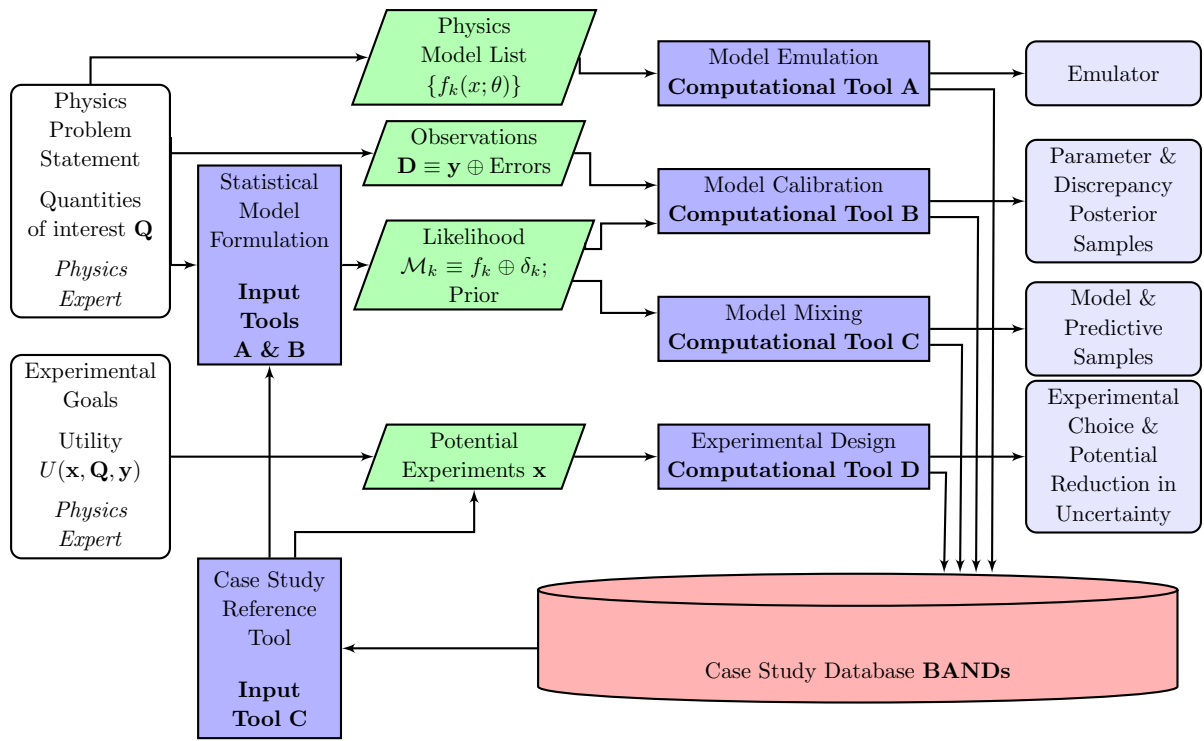
\begin{figure}[h]
\centering
\usetikzlibrary{shapes,arrows}
\usetikzlibrary{positioning}
\usetikzlibrary{arrows,calc,fit}
\tikzstyle{person} = [rectangle, draw, text width=2.25cm, text centered, rounded corners, minimum height=8em]
\tikzstyle{data} = [trapezium, trapezium left angle=70, trapezium right angle=-70,text centered,text width = 2.5cm, minimum height=1cm, minimum width=1cm, draw=black, fill=green!30]
\tikzstyle{inputtool} = [rectangle, draw,  text width=2cm, text centered, minimum height=8em, fill=blue!30]
\tikzstyle{line} = [draw, -latex']
\tikzstyle{db} =  [cylinder, shape border rotate=90, draw, text height=1cm, minimum height=1cm, minimum width=10cm,shape aspect=.1, fill=red!30]
\tikzstyle{tool} = [rectangle, draw,  text width=4cm, text centered, minimum height=1cm, fill=blue!30]
\tikzstyle{outcome} = [rectangle, draw,  text width=2.5cm, text centered, minimum height=1cm, rounded corners, fill=blue!10]

\footnotesize
\begin{tikzpicture}[thick, node distance=2.5cm, scale=.85, transform shape]
\node [person] (physicist) at (0,4) {Physics Problem Statement \\[5 pt] Quantities of interest {\bf Q}\\[5 pt] \footnotesize \emph{Physics Expert}};
\node [person] (experimentalist) at (0,0)  {Experimental Goals \\[5 pt] Utility $U({\bf x},{\bf Q},{\bf y})$\\[5 pt] \footnotesize \emph{Physics Expert}};
\node [data] at (6,6.25)  (modellist) {Physics Model List \\ $\{f_k(x;\theta)\}$ };
\node [data] at (6,4.75) (obs) { Observations\\ ${\bf D} \equiv {\bf y} \oplus {\rm Errors}$};
\node [data] at (6,3) (statmodel) {Likelihood\\ ${\mathcal M}_k \equiv f_k \oplus \delta_k$;\\ Prior};
\node [data] at (6,0) (explist) { Potential\\ Experiments {\bf x}};
\node [inputtool] at (2.8,3) (modelform) {Statistical Model Formulation\\[10pt] \footnotesize 
\bf{Input Tools A \& B} };
\node [inputtool] at (2.8,-3) (casestudy) { Case Study Reference Tool  \\[10pt] \footnotesize 
\bf{Input Tool  C}};
\node [db] at (12,-3) (database) { Case Study Database  
{\bf BANDs}};
\node [tool] at (11,6) (emu) {Model Emulation\\
{\bf Computational Tool A}
 };
\node [tool] at (11,4) (cali) {Model Calibration\\
{\bf Computational Tool  B}};
\node [tool] at (11,2) (mix) {Model Mixing\\
{\bf Computational Tool C}
 };
\node [tool] at (11,0) (exp) {Experimental Design\\
{\bf Computational Tool D}};
\node [outcome] at (16,6) (emulator) {Emulator

 };
\node [outcome] at (16,4) (paraanddisc) {Parameter \& Discrepancy  Posterior Samples
  };
\node [outcome] at (16,2) (mixingandpred) {Model \& Predictive Samples

 };
\node [outcome] at (16,0) (expchoice) {Experimental Choice \& Potential Reduction in Uncertainty }; \path [line] ($(physicist.east)-(0,-1)$) |- (obs);
\path [line] ($(physicist.north)-(0,0)$) |- (modellist);
\path [line] ($(physicist.east)-(0,-1)$) |- (modelform);
\path [line] ($(modelform.east)-(0,0)$) |- (statmodel);
\path [line] ($(experimentalist.east)-(0,0)$) |- (explist);
\path [line] (casestudy) --  ($(modelform.south)-(0,0)$) ;
\path [line] (database) --  (casestudy);
\path [line] ($(modellist.east)-(0,0)$) |- (emu);
\path [line] ($(obs.east)-(0,0)$) -- ($(obs.east)-(-0.25,0)$) |- ($(cali.west)-(0,-0.25)$);
\path [line] ($(statmodel.east)-(-0.1,-0.35)$) |- ($(statmodel.east)-(-0.35,-0.25)$) |- ($(cali.west)-(0,0.35)$);
\path [line] ($(statmodel.east)-(0.1,0.25)$) |- ($(statmodel.east)-(-0.35,0.25)$) |- ($(mix.west)-(0,-0.25)$);
\path [line] ($(explist.east)-(0,0)$) --  ($(exp.west)-(0,0)$);
\path [line] ($(emu.east)-(0,0.25)$) |- ($(emu.east)+(1,-0.25)$) -- ($(database.north)+(2.1,0)$);
\path [line] ($(cali.east)-(0,0.25)$) |- ($(cali.east)+(0.75,-0.25)$) -- ($(database.north)+(1.85,0)$);
\path [line] ($(mix.east)-(0,0.25)$) |- ($(mix.east)+(0.5,-0.25)$) -- ($(database.north)+(1.6,0)$);
\path [line] ($(exp.east)-(0,0.25)$) |- ($(exp.east)+(0.25,-0.25)$) -- ($(database.north)+(1.35,0)$);
\path [line] (emu) -- (emulator);
\path [line] (cali) -- (paraanddisc);
\path [line] (mix) -- (mixingandpred);
\path [line] (exp) -- (expchoice);
\path [line] ($(casestudy.north)+(0.4,0)$) |- ($(casestudy.north)+(0.4,0.4)$) -| ($(explist.south)$);
\end{tikzpicture}
\caption{\label{fig:flowchart} Flowchart displaying the different tools that will be incorporated in the BAND framework.}
\end{figure}

To accomplish these goals we first lay out in Sec.~\ref{sec:bayesbits} the ingredients for Bayesian inference from a dataset $\observations$ to quantities of interest (QOIs) $\qoi$ in a nuclear physics---or any---problem. These ingredients are the Bayesian prior, which encodes extrinsic information and expert opinion about the QOIs, and the likelihood, which expresses the way in which the data to be considered constrain those quantities. Within BAND, Bayesian statisticians will work with nuclear physicists on {\it prior specification} and {\it likelihood formulation}. The results will be incorporated into the software framework as ``Input Tools" A and B. These are the first steps in the flowchart for the BAND software framework, see Fig.~\ref{fig:flowchart}. 

Nuclear physicists using BAND will also specify the set of physics models from which they want to obtain a prediction. Often, evaluating these models will involve a calculation that consumes a large amount of (super)computer time for a ``forward evaluation": obtaining the observables of interest for just one instance of the model parameters. For these ``expensive" models UQ can only be accomplished in a realistic amount of time once a computationally cheap model emulator has been built. This {\it model emulation} will be accomplished by Computational Tool A. 
Emulation as a tool to reduce the computational load of inference is well covered in many references \cite{sacks1989design, santner2019design,gramacy2020surrogates}.  
We touch on it briefly in Sec.~\ref{sec:emulation}, but other than that it is not really discussed in this article. 

Once observations $\observations$ are specified by the user, BAND will combine the likelihood and prior and use emulator samples to perform \textit{model calibration}, obtaining
the posterior probability density function (``posterior" or ``posterior pdf" hereafter) for the parameters of each model (Computational Tool B). 

Even after calibration and emulation have been achieved we have still only obtained information on the individual models. Calibrating models to data, while including prior information, is a practice that is gaining increasing currency in nuclear physics. But BAND will push the field further, by taking a set of individual models, each of which have been calibrated to data, and use them to obtain a {\it model-mixed} prediction. Section~\ref{sec:bmm} discusses the general theory of model-mixed predictions, presents the standard approach of BMA, elucidates its limitations, and introduces ways to combine models that are less global, in order to leverage information on local model performance.  BMA as well as these more general BMM strategies will be implemented in Computational Tool C. 

In Sec.~\ref{sec:toy} we put the emulation, calibration, and model-mixing steps together in the context of a classical toy problem: ``the ball drop''. This (admittedly very simple) example is meant to show the kind of analysis BAND could facilitate when using several sophisticated nuclear-physics models and large sets of experimental observations. 

A major challenge in NP, as in many other advanced disciplines, is the optimal design of experiments. Not all measurements are equally useful, and beam time is expensive. The costs of running an experiment include not only the workforce, time and money invested, but also the opportunity cost of alternative measurements that were not carried out. Thus, when planning an experiment, it is important to consider which data are most likely to provide the largest information gain.  This is a highly practical field of study, with applications including engineering, biology, environmental processes, computer experiments, and psychology~\cite{chaloner1995bayesian,Liepe2013a,Ryan:2015aaa,Myung2009BayesianAO,vernon2010galaxy,Berliner2008,Cumming2009,farrow2006trade,santner2019design,currin1988bayesian,jones2016bayes}. The process of making the best selection in this regard is known as \textit{experimental design}. In order to ensure that the substantial resources necessary for modern experiments are focused on acquiring the most valuable data, both the theory uncertainty and the expected pattern of experimental errors must be considered.

BAND's model-mixed prediction is therefore important if nuclear physicists are to have  guidance on experimental design that reflects the true extent of model uncertainty. Providing such guidance will be the job of Computational Tool D. Experimental design formalism and an example of its use in a nuclear-physics context is discussed in Sec.~\ref{sec:design}.  
 
Finally, in Secs.~\ref{sec:eos}, \ref{sec:reactions}, \ref{sec:masses}, and \ref{sec:BMA_TC} we showcase different nuclear-physics problems where one or more ideas from the BAND framework have been implemented. We discuss the benefits gleaned from emulation, calibration, and model averaging in those cases. We then explain how application of the full BAND tool set will build on these initial steps towards Bayesian analyses of prominent nuclear-physics problems and yield the full benefit of using advanced statistical methods to consistently combine the insights of multiple forefront nuclear-physics models. Section~\ref{sec:conclusion} provides a summary as well as comments on topics not treated in the main text. 

Throughout the article we use a number of terms at the nuclear-physics/statistics interface. Usage frequently differs between communities, so in Table~\ref{tab:glossary} we take the opportunity to define these terms as we use them in this work.
 
\section{Finding your posterior} \label{sec:bayesbits}

At its core, a Bayesian framework seeks to obtain the probability distribution $p$ of a set of unobserved quantities of interest (QOIs) $\bf{Q}$, combining probabilistic information on beliefs about them (the prior) and on how they relate to observations $\observations$ (the likelihood). Specifically, the prior is a probability model $p(\bf{Q})$ for the QOIs, and the likelihood is a probability model $p(\observations |\bf{Q})$ for the observations given the QOIs. The output of Bayes' rule, known as the posterior, is then a probability distribution $p(\bf{Q}|\observations)$ for the QOIs given the observations \footnote{%
    We use the notation $p$ liberally for different probability notions. In particular, when we refer to the probability of a continuous quantity, $p$ should be read as a probability density function (pdf). Whether $p$ is a pdf or an integrated probability should be clear from context. For an introduction to Bayesian statistics particularly well suited to physicists, we recommend Ref.~\cite{Sivia:2006}.}.
In most modeling contexts Bayes' rule is astonishingly simple: it says that the posterior probability density of ${\bf Q}$ given $\observations$  is  proportional to the product of the prior and the likelihood:
\begin{equation}
    p({\bf Q}| \observations) \; = \; \frac{ p(\observations | {\bf Q} ) p({\bf Q}) }{p({\bf D})}  \;  = \; \frac{ p(\observations | {\bf Q} ) p({\bf Q}) }{ \int p(\observations | {\bf Q} ) p(\bf{Q}) \textrm{d} {\bf Q}}  \; \propto \; p(\observations | {\bf Q} ) p({\bf Q}). 
\label{eq:BayesRule}
\end{equation}
The functional dependence of this pdf on $\bf Q$ is given by the numerator in the middle expression. Since $\observations$ is assumed to be known, the associated denominator is just a normalization constant, whose value is not needed if one's only goal is to sample the pdf of $\bf Q$. This denominator does, however, become relevant in the context of model selection or model averaging problems. 

Prior specification and likelihood formulation are therefore the first two elements of BAND. Typically, nuclear physicists will already have an opinion as to the physics models that should be used to express a likelihood relation. The statistician's role in likelihood formulation is then to determine with clarity where the uncertainty, from both experiment and theory, comes into the NP model. How to specify priors on the  unobserved elements $\bf Q$ in a NP model is usually a much less well defined question; it is best answered through strong interactions between physicists and statisticians. We now discuss BAND's approach to prior specification and likelihood formulation before briefly describing the opportunities and challenges associated with then obtaining the posterior of the QOIs ${\bf Q}$.

\subsection{Prior specification}
\label{subsec:priors}

Specifying priors requires asking about---eliciting---prior knowledge of the quantities that are sought~\cite{oakley2002}. These could be model parameters that need to be estimated, or they could be predictions for observables that are not part of the dataset ${\bf D}$ (e.g., an interpolation or extrapolation). The statistician and the nuclear physicist need to jointly uncover expected ranges for these QOIs and any other statistical properties they wish to define for these QOIs. 

When working to encode the prior information into distributions, it is tempting to insist on the use of so-called uninformative priors with the goal of being maximally data-driven.  This approach, which
is often advocated in popular presentations of Bayesian statistics, is based on formal methods of computing the amount of information that a particular prior brings to the problem. An uninformative prior tries to minimize this information. In practice this often leads to incorrect deployment of uniform priors. 
The incorrectness can arise for several reasons~\cite{BDA}. First, a prior that is uniform in one parameterization will not be in another so ``uniform in what" is always a worthwhile question in this context. Second, uniform priors may end up being more informative than their user intends: by completely precluding certain parts of the ${\bf Q}$ domain, uniform priors can overstate what is known. But the broader problem is that uniform priors rarely reflect the actual physical prior knowledge of ${\bf Q}$.  Uninformative priors effectively lockout the logical meaning of the nuclear physics model and leave the interpretation of parameters and numerical structure to the numerical experimental results. Indeed, nuclear physicists typically have important insights into what to expect for some of the parameters or observables they seek to infer.
This prior knowledge can come from formal constraints (e.g., regarding positivity or other bounds from physical principles), from an expected size based on the physical scales in the problem, or from accumulated experience. 
By asking questions through either informal or formal elicitation, the statistician can extract some of this knowledge and build it into the priors. This facilitates the inclusion of physics information in the prior where it is warranted. Of course, checks for unwanted sensitivity to the prior should also be executed in order to catch biases in opinions that result in a misinformed prior. The prior produced by this process would be far from uninformative, and rightly so. BAND is thus built on a participatory approach to prior specification that works 
to incorporate the available and useful information about the unobserved QOIs that is not in the observations $\observations$ into the prior. 

A simple way of selecting priors in an informative way occurs by taking advantage of the fact that a prior itself has parameters. These are called hyperparameters to distinguish them from the parameters $\qoi$. The hyperparameters should be tuned to agree with the physicists' thinking while keeping with statistical principles such as prudence and parsimony. Standard distributions for parameter priors include hyperparameters that encode prior beliefs on a parameter's central value (e.g., mean) and spread (e.g., standard deviation). In practice, statisticians can gauge their NP colleagues' level of confidence in parameter ranges and other properties and advocate for distributions with hyperparameters yielding sufficiently conservative spreads or heavy tails. This type of strategy is prudent, is not computationally expensive, and can markedly increase a model's robustness. 

Informative priors are built by using other information ${\bf I}$---even if it is limited in quantity---that is relevant for the QOIs ${\bf Q}$. ${\bf I}$ should not be directly related with the information encoded in the likelihood model and the dataset ${\bf D}$.  Formally we express this relationship via repeated application of Bayes' rule:
\begin{equation}
p({\bf Q}| \observations) \propto p(\observations|{\bf Q},{\bf I}) p({\bf Q}|{\bf I}) p({\bf I}) \propto p(\observations|{\bf Q}) p({\bf Q}|{\bf I}) p({\bf I}).
\label{eq:HBayesRule}
\end{equation}
This modeling scenario is known as a hierarchical Bayesian model: the prior is not just an arbitrary set of probability distributions on each element of ${\bf Q}$, but uses other information to constrain (some of) these elements probabilistically.  The key point in the use of a hierarchical Bayesian framework is that if $p(\observations|{\bf Q},{\bf I}) = p(\observations|\qoi)$ then this is equivalent to ${\bf I}$ and the $\observations$ being independent, given $\qoi$.  In such a situation the hyperparameters that define the prior distribution $p({\bf Q}|{\bf I})$ would be estimated using ${\bf I}$. In the case that ${\bf I}={\bf D}'$ (another dataset), there is the possibility that ${\bf D}$ and ${\bf D}'$ could be analyzed simultaneously as part of a (more complicated) likelihood (see, e.g., Refs.~\cite{Barboza2014,Zhang:2015ajn}).
In that case the parameters that appear in $p({\bf Q}|{\bf I})$ 
would no longer be referred to as hyperparameters, since they would appear in the likelihood, not in the prior. 

The hierarchy that encodes the prior does not have to be complicated in order to aid the statistical determination of $\qoi$. A discussion between nuclear physicists and their statistician collaborators about the value of using a hierarchy can be initiated simply by asking what external variables or other information might be used to calibrate the knowledge the nuclear physicists want to encode in their priors.
For illustration we consider two examples of prior specification that typify NP applications.

A simple hierarchical Bayesian model can be used to aid the fitting of a polynomial of specified degree $M$. Suppose that the data to which the polynomial is fit is scaled so that the natural units of the dependent and independent variables are both of order unity~\cite{Schindler:2008fh,Wesolowski:2015fqa}. This situation is paradigmatic of attempts to extract the parameters of effective field theories (EFTs) from low-energy data. The desired quantities ${\bf Q}$ are then the model's set of parameters $\theta$, namely the coefficients $\theta \equiv  \{a_0,a_1,\ldots,a_M\}$ of the polynomial
\begin{equation}
   f(x, \theta)=a_0 + a_1 x + \ldots a_M x^M.
\end{equation} 
The likelihood relates the polynomial to the information in the dataset ${\bf D}$, which will include points where the response has been measured to have certain central values, with certain uncertainties. The key Bayesian step is to model naturalness by assuming all the coefficients $\{a_0,a_1,\ldots,a_M\}$ represent draws from a common population. 
Then the prior  on the parameters $\theta \equiv \{a_0,a_1,\ldots,a_M\}$ can be specified via hyperparameters for the mean and variance of the set of coefficients. For example, if we specify mean zero and standard deviation $\abar$ of a normal distribution,  we have:
\begin{equation}
    p(a_0,a_1,\ldots,a_M|\abar) \propto \exp\left(-\frac{a_0^2 + a_1^2 + \ldots + a_M^2}{2 \abar^2}\right).
    \label{eq:aprior}
\end{equation}
The last element in the Bayesian hierarchy would then be a prior distribution for the hyperparameter $\abar$, just as one must pick priors for any parameter. 

Another NP example of a Bayesian hierarchy arises
in the extrapolation of observables for nuclei near the driplines. In \cite{Neufcourt2018, Neufcourt2019,Neufcourt2020a,Neufcourt2020b}, separation energies are extrapolated using various Bayesian techniques, including Gaussian processes (see Sec.~\ref{sec:masses} for more discussion). For that technique, an estimation is needed for the characteristic ranges of influence of one nucleus over another in the $(Z,N)$ space. Weakly informative priors for the $Z$ and $N$ ranges-of-influence were employed, where hyperparameters for the means and variances of those priors were declared. Specifically, a Gaussian process (GP) was used to extrapolate the observable $S$ from currently known locations to a new location $(Z',N')$, with a squared-exponential kernel defining the correlation function of the GP. For two locations $(Z_1,N_1)$ and $(Z_2,N_2)$ in the nuclear landscape the correlation between the two measurements of $S$ is taken as
\begin{equation}
\textrm{Corr}(S(Z_1,N_1),S(Z_2,N_2))=  \exp\left( - \frac{1}{2} \frac{ (Z_1-Z_2)^2}{\rho_Z^2} - \frac{1}{2} \frac{ (N_1-N_2)^2}{\rho_N^2 } \right) . \label{eq:HBM-GP}
\end{equation}
Here $\rho_Z$ and $\rho_N$ are the ranges of influence. Gamma priors were chosen for their squares. A more sophisticated hierarchical Bayesian model would be to take priors for $\rho_Z$ and $\rho_N$ that depend on the mass number of the location $(Z',N')$ where we want to extrapolate, thus using a different model for each extrapolation.  Modifying $\rho_Z$ and $\rho_N$ in this manner must be carefully done to avoid violating the condition that the correlation function be positive definite, but such an adaptation allows for the inclusion of the NP knowledge that the nuclear-chart distances over which $S$ is correlated are far shorter for light nuclei than they are for heavy nuclei. A model for $\rho^2_Z$'s  and $\rho^2_N$'s mean hyperparameter that is linear in $A'=Z'+N'$ and includes an additive error term captures this belief and admits uncertainty about it. This means two new hyperparameters will need to be determined: the slope of the linear model with respect to $A'$, and the noise level there. An even more sophisticated hierarchical Bayesian model that has four hyperparameters rather than two might take $\rho^2_N$ to have a different slope and a different noise level than $\rho^2_Z$, because the valley of stability is longer than it is wide. These ways of defining the prior distribution of $\rho_Z$ and $\rho_N$ would produce a conditional GP for $S$, where the range of influence is uncertain and depends on the extrapolation location of interest. But it is unlikely that any nuclear physicist would just say ``Hey, let's write down a conditional Gaussian process for this correlation matrix, which depends on individual extrapolation conditions". The hierarchy enables the organized and clear incorporation of known physics in the probabilistic model.  The BAND-driven collaboration is designed to match insight in nuclear physics with statistical tools exactly as done in this example.

To summarize, the task of picking priors is nontrivial, yet priors can have a fundamental influence on the statistical analysis. Informative priors can be useful and should not be shunned. Overstating what we know, by picking priors that are excessively informative, can lead to problems like credibility intervals for the QOIs that are too narrow. Understating what we know is also a mistake, and is liable to lead to credibility intervals that are too wide.

\subsection{Likelihood formulation}
\label{subsec:likelihood}

We now define our notational convention for setting up likelihood models of the form most commonly used in nuclear physics. A deterministic physics model (i.e., one with no randomness) that nominally explains observable $y$ (e.g., cross section, masses) from an input $x$ (e.g., kinematics, proton and neutron  numbers), will take the functional form $y=f(x,\theta)$, where $\theta$ represents parameters which may need to be estimated. In a set of observations $\genobsset \equiv \{y_i:i=1,\ldots,n\}$ at points $\design \equiv \{x_i:i=1,\ldots,n\}$ there will be disagreement with the physics model. Because of this we write the relationship between those observations and the physics model as $y=f(x,\theta)+\text{error}$. This model for the observations then includes both a physics model, which may depend on unobserved parameters, and a statistical model for the error term. 

The familiar so-called $\chi^2$ formulation follows when the statistical model assumes that the error at each experimental measurement point $i$ is independent and normally distributed~\footnote{Other distributions can certainly be used, but we have assumed normally distributed uncertainties here since that case is the one with which readers are likely to be most familiar.} with mean $0$ and variance $\sigma_i^2$, namely
\begin{equation}
p(\observations | \theta,\{\sigma_i^2\} ) \propto \exp\left( - \frac{1}{2} \sum_{i=1}^n \frac{ \bigl(y_i - f(x_i,\theta)\bigr)^2}{\sigma_i^2 } \right). \label{eq:standardL}
\end{equation}
Throughout the article, $\observations$ represents the list of couples $(x_1,y_1),\ldots,(x_n,y_n)$, and so $\observations \equiv \{\design = (x_1,\ldots,x_n),\genobsset = (y_1,\ldots,y_n)\}$ includes  both the input choices and the experimental observations~\footnote{Strictly speaking, this definition of $\observations$ means that $\design$ has been moved to the other side of the conditional in (\ref{eq:standardL}) because we presume the $\qoi$'s we are trying to infer do not depend on where we make the observations.}. 
The physics model $f$ may depend on unknown parameters $\theta$; the intensity of the point-to-point error is also sometimes unknown. In Eq.~(\ref{eq:standardL}) we have denoted explicitly that the pdf depends on this intensity of errors $\{\sigma_i^2\}$.  A subtle point is that this expression implicitly is conditional on a physics model $f$.  The suppression of obvious conditionals is common in Bayesian statistics: it prevents page-long expressions and emphasizes the key data and parameters. This implicit conditioning on the physics model will become important later when we turn our attention to emulation and mixing, but it remains implicit for now. Conversely, in later applications some dependencies that are explicit on the right of the conditional here become implicit.

Heterogeneous datasets often appear in the likelihood. In such cases, the dataset $\observations$ can be divided into $n_{\rm cl}$ classes of observations $\observations_1,\dots,\observations_{n_{\rm cl}}$. The data classes may contain rather different numbers of observations and the level of precision may vary widely between classes too. For instance, the data class $\observations_1$ may represent $100$ binding energies, the  data class $\observations_2$ may represent $10$ charge radii, and so on. Breaking up the data into different data classes facilitates using different covariance forms for each class, which has the effect of introducing relative weights for each class into the likelihood, so that one can avoid a situation in which one data type dominates because it is either very numerous or very precise \cite{Birge1932,Dobaczewski2014}.

Notice that in Eq.~(\ref{eq:standardL}) we have deliberately not stated whether the noise term $\sigma_i$ comes from experimental noise and/or imperfections in the theoretical model. If the form \eqref{eq:standardL} is used in the presence of model imperfections,  the  assumption stated above is implicitly adopted for theoretical errors as well. 

But theoretical errors are typically highly correlated. When model imperfections are a significant contributor to the overall uncertainties, a likelihood that uses a non-diagonal covariance matrix
may be a better choice. For example, in the polynomial-coefficient parameter estimation problem discussed in the previous section, we can estimate the coefficients in the $k$th-order polynomial while treating the term of $O(x^{k+1})$ as a model imperfection. If we then marginalize over the coefficient $a_{k+1}$ using the ``naturalness" information in the prior (\ref{eq:aprior}) we obtain a modified likelihood~\cite{Schindler:2008fh,Wesolowski:2018lzj}:
\begin{equation}
    p(\observations | \theta,\Sigma ) \propto \exp\left( - \frac{1}{2} \sum_{i,j=1}^n \bigl(y_i - f(x_i,\theta)\bigr) \Sigma^{-1}_{ij} \bigl(y_j -  f(x_j,\theta)\bigr)  \right). \label{eq:modifiedL}
\end{equation}
Here the matrix $\Sigma$ can  be expressed as $\Sigma=\Sigma_{\rm exp} + \Sigma_{\rm th}$, where $\Sigma_{\rm exp}$ is the diagonal covariance matrix used in Eq.~(\ref{eq:standardL}) above:
\begin{equation}
    \Sigma_{\rm exp} \equiv \mbox{diag} (\sigma_i^2:i=1,\ldots,n),
\end{equation}
while the piece of $\Sigma$ associated with the theory error encodes a high degree of correlation:
\begin{equation}
     \Sigma_{\rm th, ij}=\sigma_a^2 x_i^{k+1} x_j^{k+1}.
\end{equation}
Similarly, if the point-to-point (``statistical") and systematic uncertainties in an experiment are accurately characterized and well explained in the publication detailing the observations, then it is straightforward to write down a likelihood with a non-diagonal covariance matrix that accommodates components of the experimental uncertainties that are not independent (see, e.g., Ref.~\cite{Stump:2001gu}). 

All such generalizations, where observations $(x,y)$ are modeled as functions of unobserved quantities $\theta$, and where we incorporate probability modeling for a random error of possibly unknown intensity, yield a likelihood derived from a statistical model $y=f(x,\theta)+\text{error}$. These likelihoods encode the statement ``This is how likely we think it would be to observe what we see $y$, under conditions $x$, based on the model function $f$ that depends on parameters $\theta$, and based on an error intensity $\sigma$". Equation~\eqref{eq:standardL} provides a particularly simple example of this kind of statistical model and it is used very often. 

But, in fact, the likelihood formulation $y=f(x,\theta) + \text{error}$ does not mandate that the operand ``$+$" be interpreted as an additive error. For example, it can be formulated so that the function $f$ itself is a random distribution (i.e., not a deterministic model) where the values $x$ are used to define the distribution's parameters. A specific instance of this is when a Gaussian process (GP) is used to directly interpolate or extrapolate to QOIs. What all likelihood formulations have in common in the Bayesian context is that, when they are combined with a suitable prior according to \eqref{eq:BayesRule}, they (i) provide a principled solution to the inverse problem of estimating QOIs by introducing priors for them (and for $\sigma$, if needed); and (ii) use probability models.

Finally, we reiterate that the ``error" should account for imperfections in both the model and the experiment. It is advisable to consider a component of the error which we call a discrepancy and that represents model imperfections: the $\delta(x)$ that appears in the likelihood  (\ref{eq:Likelihoodwithdiscrepancy}) is an example of such a term. This error component depends on observables and experimental conditions, and is often correlated in the domain of $x$ values. 

\subsection{Together again: combining the prior and the likelihood and how to deal with what you get} 

Once prior and likelihood models/distributions have been agreed upon, it typically becomes a conceptually trivial matter to write down posteriors for the QOIs given the data and these agreed-upon models, see Eq.~(\ref{eq:BayesRule}). For illustration, in this article there are also examples of how to extrapolate experimentally inaccessible values $\tilde{y}$ for experimentally inaccessible conditions $\tilde{x}$ (see Sec.~\ref{sec:masses}). The method for this is to use Bayesian prediction, where the likelihood distribution of $y$ given $x$ under parameters $\theta$, applied to the range of values of interest $\tilde{x}, \tilde{y}$, is integrated against the posterior distribution of parameters $\theta$:
\begin{equation}
p(\tilde{y}|\tilde{x},\observations)=\int_{\theta}p(\tilde{y}|\tilde{x},\theta,\observations)p(\theta|\tilde{x},\observations) \textrm{d}\theta .
\label{eq:Bayesprediction}
\end{equation}
The result of the integration is known as the ``posterior predictive distribution".
For the typical scenario in NP the data influences the distribution for $\tilde{y}$ explicitly only through the parameters, and the posterior distribution of $\theta$ is thought to be independent of the hypothetical experimental conditions $\tilde{x}$, in which case Eq.~\eqref{eq:Bayesprediction} simplifies to
\begin{equation}
p(\tilde{y}|\tilde{x},\observations)=\int_{\theta}p(\tilde{y}|\tilde{x},\theta) p(\theta|\observations) \,\textrm{d}\theta .
\label{eq:Bayespredictionsimpler}
\end{equation}

The challenge then becomes understanding how posteriors like Eqs.~\eqref{eq:BayesRule}, \eqref{eq:HBayesRule}, and \eqref{eq:Bayespredictionsimpler} depend on all the variables and parameters involved. Typically, as soon as there is more than one unknown parameter, and unless priors are set up in extremely specific (and not necessarily realistic) ways, the behaviors of the resulting posterior parameter and predictive distributions cannot be obtained analytically. Means, modes, variances, etc., cannot usually be computed explicitly. One then resorts to mathematical simulations (e.g., Markov Chain Monte Carlo (MCMC) sampling) to extract information about these distributions. But our concern here is not with the specific implementation used to obtain the posterior; instead we seek to illuminate the structure and benefits of combining a Bayesian statistical model with a physics model in order to improve the inference of the physics of interest. 

\section{Bayesian inference for multiple models} \label{sec:bmm}

In this section we discuss the challenge of combining the insights from a number of individual physics models to produce inference endowed with the physics models' collective wisdom. Section~\ref{subsec:theory} provides the general setup for this problem, and introduces the crucial distinction between ${\cal M}$-closed and ${\cal M}$-open settings. Section~\ref{subsec:BMA} describes the standard Bayesian solution: Bayesian Model Averaging (BMA); we then explain why BMA can only resolve the challenge in the ${\cal M}$-closed context. Section~\ref{subsec:BMMgeneralities} then articulates paths to generalize BMA to a more sophisticated Bayesian Model Mixing (BMM), wherein we combine information from different models in a more textured way than BMA accomplishes. We end with Sec.~\ref{subsec:stat_example}, which gives an example where BMM improves upon BMA by leveraging information on the local performance of two different models across the input domain.

\subsection{Bayesian inference in the multi-model setting}

\label{subsec:theory}

Recall that our generic setup is that we have observations $\observations$ consisting of pairs of inputs and outputs $(x_1,y_1),\ldots,(x_n,y_n)$ and 
want to, from these, predict quantities of interest $\qoi$, which could be parameters, or interpolations or extrapolations, or even some totally new observable.
In this section we further suppose we have several  physics models $f_k$  ($k = 1,\ldots,K)$ that are purported to be a mapping from an  $x$ to a $y$.  Each physics model takes in an input setting $x \in \mathcal{X}$ and a parameter setting $\theta_k \in \Theta_k$. The $k$th physics model is represented  by $f_k(x,\theta_k)$, which should be considered a deterministic prediction of the observable at $x$ once the model $k$ and parameters $\theta_k$ are specified.  One can build a  model $\mathcal{M}_k$ for  observables by combining a physics model with an error term $\varepsilon$ that represents all uncertainties (systematic, statistical, computational):
\begin{equation}
\mathcal{M}_k : y_i = f_k(x_i,\theta_k) + \varepsilon_{i,k} \label{eq:stand_model}
\end{equation}
Usually, $\varepsilon_{i,k}$---the error of the $i$th observation in the $k$th model---is decomposed into a stochastic term modeling systematic discrepancy and an independent term \cite{KoH,higdon2004combining}.  Note that the error does not always have to be an additive form, but we have displayed it as such for simplicity. Moreover, as written above, $\varepsilon_{i,k}$ depends on the physics model as well as on (hyper)parameters describing the statistical model, but this notation is suppressed as the dependence involves complex factors \cite{plumlee2017bayesian}.

While different physics models may have different parameters, inference on multiple models involves dealing with a canonical parameter space $\Theta$ that spans all models of interest. We assume that for each $k$ in $\{1,\ldots,K\}$, the model-specific parameter space $\Theta_k$ can be mapped to $\Theta$ via some (possibly non-invertible) map $\mathcal{T}_k: {\Theta}_k \mapsto \Theta$. 
After transformation, we say the parameters are in the canonical parameter space, and simply write our canonical parameter as $\theta \in \Theta$ since $\Theta$ is common to all models after the application of $\mathcal{T}_k.$  We can think of this overall parameter space $\Theta$ as the union of the individual (transformed) model-specific parameters arising out of each model.  For notational simplicity, the $\mathcal{T}_k$ function will be suppressed throughout this article, meaning $\theta$ is understood as $\mathcal{T}_k(\theta_k)$ when appropriate. 

Our goal is to conduct inference on the values of $\theta$ as well as the error term $\varepsilon_{i,k}$ for each model using Bayesian inference. Three conceptual settings have been identified  (see, e.g., \cite{Bernardo94}) where Bayesian inference on multiple models is applied: $\mathcal{M}-$closed, $\mathcal{M}-$open, and $\mathcal{M}-$complete.  These three settings were originally motivated in the context of statistical model building.  In the $\mathcal{M}-$closed case, one has `closed off' the need to introduce new models as it is known that the {\em perfect model} that represents the physical reality  must be within the set of models being considered. Therefore, as data become more numerous and/or precise in the $\mathcal{M}-$closed case, that perfect model will become increasingly more likely, ultimately to the exclusion of all other models under consideration.  
In the $\mathcal{M}-$open case, one is open to introducing new  models since the perfect model is not known.  In the $\mathcal{M}-$complete case, we have decided that while we might introduce new models for the sake of accuracy, we would like to maintain inference on those in our original model set.  We will not discuss this last case further.

The key distinction for inference in nuclear physics is between $\mathcal{M}-$closed, when the set of models is expected to include the perfect one, and $\mathcal{M}-$open, when we know that the set of models does not include the perfect one.  We briefly outline the standard statistical solution for the $\mathcal{M}-$closed setting in the next section before moving on to describing some potential approaches for the $\mathcal{M}-$open setting that is more  interesting in the context of the BAND framework.

\subsection{Bayesian model averaging and the \texorpdfstring{$\mathcal{M}$}{M}-closed assumption}

\label{subsec:BMA}

Historically, mixing together different statistical models has been done through Bayesian model averaging (BMA) \cite{BMA,Was00}.  BMA has been broadly applied in many areas of research including the physical and biological sciences, medicine, epidemiology, and  political and social sciences. For a recent survey of BMA applications, we refer to 
\cite{Fragoso2018}.    BMA is a  framework where several competing (or alternative) models $\mathcal{M}_1, \dots, \mathcal{M}_K$ are available. The BMA posterior density $p(\qoi|\observations)$ corresponds to the linear combination of the posterior densities of the individual models:
\begin{equation} \label{eqn:posteriorBMA}
    p(\qoi|\observations) = \sum_{k=1}^K p(\qoi|\observations,\mathcal{M}_k)\, p(\mathcal{M}_k|\observations) .
\end{equation}
If we pull through the typical inference, we can compute the first term $p(\qoi|\observations,\mathcal{M}_k)$ by 
\begin{equation}\label{pofm1}
p(\qoi|\observations,\mathcal{M}_k)=\int_{\Theta} p(\qoi|\observations,\mathcal{M}_k,\theta) p(\theta|\observations,\mathcal{M}_k) \mathrm{d} \theta.
\end{equation}
The second term in Eq.~(\ref{eqn:posteriorBMA}), $p(\mathcal{M}_k|\observations)$, represents the posterior probability that the  model $k$ is correct. It can be computed as
\begin{equation}\label{pofm}
p(\mathcal{M}_k|\observations) = \frac{p(\observations|\mathcal{M}_k)p(\mathcal{M}_k)}{\sum_{k=1}^K p(\observations|\mathcal{M}_k)p(\mathcal{M}_k)}
\end{equation}
where
\begin{equation}
p(\observations|\mathcal{M}_k)= \int_{\Theta} p(\observations|\mathcal{M}_k,\theta) p(\theta|\mathcal{M}_k) \mathrm{d} \theta.
\end{equation}
The BMA posterior (\ref{eqn:posteriorBMA}) for $\qoi$ can then be obtained by using (\ref{pofm1}) and (\ref{pofm}).

The posterior probability of model $k$ being correct, 
$p(\mathcal{M}_k|\observations)$,
accounts for the common physics assumptions or phenomenological properties being studied that may span many of these models.  But this framing works by choosing a single model that is dominant over the entire model space.    If a perfect model is explicitly considered, that is, if some $\mathcal{M}_k$ is correct, the corresponding term should dominate the sum in  (\ref{eqn:posteriorBMA}). However, generic BMA can lead to misleading results when a perfect model is not included.  One illustration is presented in  Sec.~\ref{subsec:stat_example}.  No nuclear physics models have access to an exact representation of reality; one only hopes some are usefully close to it. It is to be noted that while  using an $\mathcal{M}-$closed approach  may be problematic in many nuclear physics applications, there are nuclear physics cases when BMA can be useful~\cite{Jay2020}. 

But, more generally, to be useful for nuclear physics, Bayesian inference methods should account for the relative performance of models among the different observables.  Some early efforts in this direction include \cite{kejzlar2019bayesian,Kejzlar2020} which consider multiple models which do not live on a common domain, resulting in some models being useful for prediction in certain physical regimes but not others. 

\subsection{Using Bayesian model mixing to open the model space}

\label{subsec:BMMgeneralities}

Suppose then, that no models are exactly correct through the domain of interest. To conceptualize this situation we introduce 
 notation for the physical process $f_\star(\cdot, \theta)$, which gives the perfect (or oracle) model. 
That model's predictions are related to the experimental observations by:
\begin{equation}
 y_i = f_\star(x_i,\theta) + \varepsilon_{i,\star}, \label{eq:true_model}
\end{equation}
where the set of $\varepsilon_{i,\star}$'s represent the error between the perfect model and  imperfect observations. Equation (\ref{eq:true_model}) is introduced purely for conceptual purposes. It is not practical because only an oracle has access to $f_\star(\cdot, \theta)$.  Someone who knows $f_\star$ because they have direct access to the underlying reality of the universe would likely not be bothered with statistical inference---or with the scientific process at all.  By presuming the $\mathcal{M}-$open scenario we invite the possibility that there is no $k$ for which $f_\star(\cdot, \theta)$ is equivalent to $f_k(\cdot,\theta)$.   The challenge is if that is true it breaks the statistical modeling principles that undergird the effectiveness of BMA as an inferential strategy. 

The generalized alternative framework we now present does not attempt to weight models based on their performance across the entire input space.  We say that such a generalized framework is an example of Bayesian model mixing (BMM).  Our approach has connections to existing statistical literature such as \cite{goldstein2009reified} in addition to the single-model frameworks of \cite{KoH} and \cite{higdon2004combining}.  Our objective is to establish different distributional assumptions beyond the assumption that any one model is perfect throughout the input space.    We do this by constructing a model $\mathcal{M}_\dagger$ that combines the physics models to inform on the observations:
\begin{equation}
\mathcal{M}_\dagger: y_i = f_\dagger(x_i,\theta) + \varepsilon_{\dagger,i},  \text{ where }  f_\dagger(\cdot, \theta) \text{ is formed by combining }f_1(\cdot,\theta),\ldots,f_K(\cdot,\theta). 
\label{eq:completeModel}
\end{equation}
The \emph{supermodel}  $f_\dagger$ is built to contain the collective wisdom of all existing models (this model was also termed reified in  Ref.~\cite{goldstein2009reified}).
One possible way to combine the models is BMA, where $f_\dagger(\cdot, \theta)$ has a prior distribution that is a point mass at each of $\{f_k(\cdot, \theta):k=1,\ldots,K\}$ that holds universally throughout the domain of interest. In BMM, we open up the possibility to combine the $K$ models in more sophisticated ways.   By mixing, one can form many potential inferences about $f_\dagger$, and---we hope---produce inferences using $f_\dagger$ that more closely resemble inferences produced by the oracle using $f_\star$.

The mixing approach would then give $p(\qoi|\observations) = p(\qoi|\observations,\mathcal{M}_\dagger).$  BMA is thus a particular special case of the BMM approach.  The key to the BAND BMM framework is that $\mathcal{M}_\dagger$ accounts for underlying information present in the individual models. In the next subsection we present an example where such an $\mathcal{M}_\dagger$ is constructed in a way that takes into account the different places in the input domain $\mathcal{X}$ in which each of them is more accurate.  

\subsection{A tale of two models: contrasting BMA with BMM} \label{subsec:stat_example}

Let us discuss a brief statistical example to unpack the sometimes subtle difference between BMA and BMM.  This should not be considered a general assessment of the approaches, but instead an example to ground the concepts. For simplicity of presentation, we assume that we have two physics models: $f_1(\cdot,\theta)$ and $f_2(\cdot,\theta)$. We want to combine these two models to produce a model $f_\dagger$ that is as close to the perfect model $f_\star$ as possible. Since perfection is not attainable we distinguish between $f_\star$, which we continue to use as a {\it gedankenmodel}, and $f_\dagger$ and try only to build the latter. 

The first of the two models being mixed, $f_1$, is an imperfect model everywhere. Conceptually we imagine that, for all values of $x\in \lbrace x_1,\ldots,x_n\rbrace$, $f_1$ differs from $f_\star$ by a stochastic discrepancy a priori normally distributed with mean zero and some moderate variance. In contrast the second model, $f_2$, is such that there is a single observation, say the one at the first point $x_1$, for which $f_2(x_1,\theta) - f_\star(x_1,\theta)$ is potentially very large, i.e., here we think that the stochastic discrepancy is normally distributed with mean zero and an extremely large variance. But everywhere else the model is essentially perfect.
We convert this information into Bayesian inference for $f_\dagger$ by saying that $f_1(x_i,\theta)$ given $f_\dagger(x_i,\theta)$ is normally distributed with mean $f_\dagger(x_i,\theta)$ and variance $v_1$.   And that $f_2(x_1,\theta)$ given $f_\dagger(x_1,\theta)$ is normally distributed with mean $f_\dagger(x_1,\theta)$ and variance $v_2 \gg v_1$, while, for $j = 2,\ldots,n$, we have $f_2(x_j,\theta) = f_\dagger(x_j,\theta)$.  

A BMA approach that acknowledges these model discrepancies expands the observed variance by the model error variance.  We will assume each model has the same prior probability of being correct and the prior $p(\theta)$ on $\theta$ is given such that $p(\mathcal{M}_1,\theta) = p(\mathcal{M}_2,\theta) = \frac{1}{2}p(\theta)$.  In terms of a posterior on the parameters, see (\ref{eqn:posteriorBMA}), this implies that
\begin{align}\label{simplemodel}
p_{\rm BMA}(\theta | \observations) \propto &\ p(\theta)\left[\prod_{i=1}^n\frac{1}{\sqrt{\sigma_i^2 + v_1}}  \exp \left(- \frac{1}{2} \frac{(y_i - f_1(x_i,\theta))^2}{\sigma_i^2 + v_1} \right)\right. \\\nonumber
+&\left.\frac{1}{\sqrt{\sigma_1^2 + v_2}}  \exp \left(- \frac{1}{2}  \frac{(y_1 - f_2(x_1,\theta))^2}{\sigma_1^2 + v_2}  \right) \prod_{i=2}^n\frac{1}{\sigma_i}  \exp\left(- \frac{1}{2}  \frac{(y_i - f_2(x_i,\theta))^2}{\sigma_i^2}   \right) \right] 
.
\end{align}
As mentioned previously, the BMA approach presumes that one model is correct throughout the entire domain of interest. If $v_2$ is truly extremely large, the BMA formalism will implement this presumption in the most extreme way possible. The spectacular failure of the second model at the first data point causes it to lose badly to the first model which just manages to be mediocre everywhere.  That is, the expression for the posterior when $v_2 \rightarrow \infty$ becomes
\begin{equation}
p_{\text{BMA}}(\theta|\observations) \propto \exp \left(- \frac{1}{2} \sum_{i=1}^n \frac{(y_i - f_1(x_i,\theta))^2}{\sigma_i^2 + v_1} \right) p(\theta).
\end{equation}
The model  $f_2$ has no role in the BMA posterior because the BMA weights consider only the overall performance of the model over the entire  domain of interest! But it seems unduly wasteful to discard the entirety of $f_2$ because it performs poorly in one small subset of the  domain of interest.

Now we consider a BMM approach where we do not presume a single model is correct throughout the entire input space.  
One potential BMM approach obtains the distribution of $f_\dagger(x,\theta)$ by using standard Bayesian updating formulae to combine the probability distributions
of $f_1(x,\theta)$ and $f_2(x,\theta)$ given $f_\dagger(x,\theta)$
with a Normally distributed prior on $f_\dagger$ having variance $v_\dagger$. Taking $v_{\dagger} \rightarrow \infty$, we have
\begin{equation}
\label{eqn:fstarexample34}
f_\dagger (x_i, \theta) \text{ is distributed as} \begin{cases}
\mathcal{N} \left(  \frac{v_2 f_1(x_i,\theta) + v_1 f_2 (x_i,\theta)}{ v_1  + v_2}, \frac{v_1v_2}{ v_1  + v_2} \right)& \text{ if }  i = 1 \\
f_2(x_i, \theta) & \text{ if } i = 2,\ldots,n.
 \end{cases}
\end{equation}
This seems to use our inference on both $f_1$ and $f_2$ in an effective way.  Pulling this into a posterior, we get that at $v_2 \rightarrow \infty$ 
\begin{equation}
p_{\text{BMM}}(\theta|\observations) \propto \exp \left(- \frac{1}{2}  \frac{(y_1 - f_1(x_1,\theta))^2}{\sigma_1^2 + v_1} -\frac{1}{2}   \sum_{i=2}^n \frac{(y_i - f_2(x_i,\theta))^2}{\sigma_i^2} \right)  p(\theta).
\end{equation}
Now both models are being used in their respective strong areas: the model $f_2$ is ignored only at a single point $x_1$ where it is very wrong and $f_1$ is  ignored everywhere that $f_2$ provides a perfect result. 

This example illustrates nicely that BMM can be a more effective tool for combining models than BMA. Although the example is simple we believe the concept it represents has wide applicability  in NP applications  where the models we want to mix perform well in different regions of the domain of interest.

\section{An illustration: using BAND framework tools to analyze a toy problem} \label{sec:illu_examp}
\label{sec:toy}


We now outline a toy example that spans the emulation, calibration and model-mixing components of the BAND framework. The experimental design component of BAND is discussed in Sec.~\ref{sec:design}.  To facilitate the discussion, we will mostly make use of a basic GP toolset. GPs are a popular default modeling choice for a few reasons, including: their prior-on-functions interpretation, the smooth, continuous and differentiable emulations they can provide, and their effectiveness when emulating sparsely observed functions. 
We will outline a basic approach to emulating, calibrating and mixing these models as would be desired in a real nuclear physics investigation---keeping in mind that the BAND framework aims to enable multiple tools (i.e., a library of emulators, model mixing methods, and experimental design algorithms) to be used in an inter-operable and consistent manner. The simplified toy example we outline in this section can be further explored in the R script file located in the BAND GitHub repository \cite{BANDGithub}.


\subsection{The toy model}

In line with the notation established in the previous section we take a toy model, $\mathcal{M}_k$, to involve a physics model $f_k(x,\theta)$ that depends on a single input $x$ and a parameter $\theta.$  
Given this known $\theta$, we can compute $f_k(x,\theta)\equiv f_k(x)$ at a selection of $m_k$ settings of the input, ${\bf x}_k \equiv (x_1,\ldots,x_{m_k})$ giving 
model outputs ${\bf f}_k \equiv (f_k(x_1),\ldots,f_k(x_{m_k})).$  

A popular toy model we will use to outline the BAND framework arises in the so-called ball drop experiment~\cite{Higdon:etal:2010}.  In this experiment, a large ball is dropped from a tower, and its height is recorded at discrete time points until it hits the ground.  The input, $x$, is time and the observable of interest, $y$, is the ball height.  We will eventually consider two particular toy models for this physical process:
\begin{itemize}
\item[$\mathcal{M}_1$:] A model for ball height that ignores atmospheric drag due to air resistance.  The physics model, $f_1,$ depends on a single parameter $\theta=g$, the acceleration due to gravity. 
\item[$\mathcal{M}_2$:] A model for ball height that includes a quadratic component for atmospheric drag due to air resistance.  The physics model, $f_2,$ depends on two parameters, $\theta=(g,\gamma)$ where $\gamma$ is a drag coefficient.
\end{itemize}
The physics of both models are outlined in \cite{Taylor:2005}, and our toy problem will involve dropping a $0.1$~m diameter ball weighing $1$~kg.



\subsection{Emulation}

\label{sec:emulation}

We start with our simpler model, $\mathcal{M}_1$, which will only be an accurate description of the physics when the effect of drag can be ignored.  For simplicity, we simulate our observables directly from $\mathcal{M}_1$ at the ``true'' gravity parameter $g=9.8$~m/s$^2$.

Our first task of interest is to predict, or {\em emulate} \cite{sacks1989design,currin1991bayesian,santner2019design,vicario2016meta,gramacy2016lagp,iooss2019advanced,katzfuss2020scaled,plumlee2020composite,konig2020eigenvector}, our physics theory $f_1(x)$ at arbitrary input(s) $\tilde{x},$ which were not made available to us directly from the output of physics model $\mathcal{M}_1.$  As outlined in Sec.~\ref{Intro}, emulation is a probabilistic technique that provides a computationally cheap surrogate for a model when the model can only be evaluated at a sparse selection of input settings. This allows one to explore questions of interest when evaluation of the model is limited due to computational constraints. To perform this emulation, a prior distribution, $p_{\rm Emulate}({\bf f}_1\vert {\bf x}_1,\phi),$ describes the statistical emulator to be used. 
Here, $\phi$ refers to {\em nuisance} parameters that are necessary for the statistical emulator, but are not directly physics parameters of interest. Without loss of generality, we will drop $\phi$ from the notation unless required for clarity.

Emulation is then the process of probabilistically recovering the rest of $f_1$ using only the observed model runs $({\bf f}_1,{\bf x}_1),$ and the prior distribution $p_{\rm Emulate}$. Suppose we want to emulate $f_1$ at a point $\tilde{x}$. This task is performed via the {\em  posterior predictive distribution},  which is obtained by integrating over the emulator nuisance parameters $\phi$:
\begin{equation}
p_{\rm Emulate}(f_1(\tilde{x})\vert \tilde{x},{\bf f}_1,{\bf x}_1):=\int_\phi p_{\rm Emulate}(f_1(\tilde{x})\vert \tilde{x},{\bf f}_1,{\bf x}_1,\phi)p(\phi \vert {\bf f}_1,{\bf x}_1)d\phi.
\end{equation}
A key ingredient of the posterior predictive distribution is the first term of the integrand, $p_{\rm Emulate}(f_1(\tilde{x})\vert\tilde{x},{\bf f}_1,{\bf x}_1,\phi),$ which encodes how the observed function values ${\bf f}_1$ are used to probabilistically extrapolate our function's behavior at new input setting $\tilde{x}.$  Meanwhile, the second term, $p(\phi\vert{\bf f}_1,{\bf x}_1)$ encodes the information learned about our function from the finite outputs ${\bf f}_1$, such as the function's smoothness or differentiability.
Note then that this Bayesian solution describes an entire emulation pdf. A typical point estimate---i.e., the thing we might quote for ``the number" given by the emulator---would be the mean of the posterior predictive,
\begin{equation}
E[f_1(\tilde{x})\vert \tilde{x},{\bf f}_1,{\bf x}_1]=\int_{f_1(\tilde{x})}f_1(\tilde{x})p_{\rm Emulate}(f_1(\tilde{x})\vert \tilde{x},{\bf f}_1,{\bf x}_1)df_1(\tilde{x}).
\end{equation}
But although this provides us with a ``the number", it is important to note that the posterior predictive distribution is just that: a distribution, and as such the emulator comes with an emulator uncertainty that is encoded in the spread and other properties of that distribution. The development of GP emulators for this problem is thoroughly discussed in \cite{santner2019design,gramacy2020surrogates}.

\subsection{Calibration}
 In statistical calibration, we expand on the emulation described above by removing the assumption that we know $\theta$ while also introducing
a model discrepancy term, $\delta(x)$ that allows for the possibility of model misspecification.
Calibration is a powerful technique because it allows one to combine sparse observables with sparse emulator outputs to perform inference and predictions.  If emulation is not required, the extension of Eq.~(\ref{eq:standardL}) to include the discrepancy term, $\delta(x)$, is
\begin{equation}
p(\observations  \vert f_k,\theta,
\delta,\lbrace\sigma_i^2\rbrace) \propto \exp\left( - \frac{1}{2} \sum_{i=1}^n \frac{ (y_i - f_k(x_i,\theta)-\delta(x_i))^2}{\sigma_i^2 } \right).
\label{eq:Likelihoodwithdiscrepancy}
\end{equation}

In the more common case where emulation is needed, we choose to run our physics model at only $m_k$ settings because every such run is costly in time, money, or some other thing we care about. Each selected ``setting'' corresponds to a simultaneous choice of inputs and calibration parameters, and we notate those settings hereafter as ${\bf x}_k\equiv (x_1,\ldots,x_{m_k})$  and $\boldsymbol{\theta}_k \equiv (\theta_1,\ldots,\theta_{m_k})$.  Outputs from our physics model $\mathcal{M}_k$ then comprise ${\bf f}_k \equiv \left(f_k(x_1,\theta_1),\ldots,f_k(x_{m_k},\theta_{m_k})\right).$  
Let ${\bf C}_k=\lbrace {\bf f}_k,{\bf x}_k,\boldsymbol{\theta}_k\rbrace$.
Calibration assumes there are two sparse sources of information: $n$ real-world observables, ${\bf y}$, and $m_k$ outputs from a physics model of interest, ${\bf f}_k$.  
These two sources of data are then combined in a statistical model, $p_{\rm Emulate}({\bf y},{\bf f}_k\vert {\bf x}, {\bf x}_k,\boldsymbol{\theta}_k,\theta,\delta)$ that connects the observations with model outputs  conditional on knowing both the calibration parameter setting that best aligns with reality, and the model discrepancy term, $\delta,$ that accounts for infidelity between the physics model and reality.  Note that this means the ${\bf C}_k$ is divided in $p_{\rm Emulate}$---as $\observations$ was in Eq.~(\ref{eq:standardL})---since the model treats the $\boldsymbol{\theta}_k,{\bf x}_k$ as fixed and known in order to emulate the ${\bf f}_k$ and ${\bf y}.$  

Calibration then allows two distributions of interest to be calculated. First, there is the {\em posterior distribution} for $\theta$ and $\delta$. By Bayes' theorem (\ref{eq:BayesRule}) that is:
\begin{equation}
p_{\rm Calibrate}(\theta,\delta\vert {\bf C}_k,{\bf D}) \propto p_{\rm Emulate}({\bf y},{\bf f}_k\vert {\bf x},{\bf x}_k,\boldsymbol{\theta}_k,\theta,\delta) p(\theta)p(\delta).
\end{equation}
Here, we see that the posterior distribution encodes how much information was learned about the unknown calibration parameter setting $\theta$ that aligns with the observables ${\bf y},$ and it also encodes what was learned about potentially unaccounted for physics, $\delta,$ in our function $f_k.$  Note that this is accomplished using only a finite sample of observables and model outputs.

Second, there is the posterior predictive distribution which, as in Eq.~(\ref{eq:Bayesprediction}), can be found by marginalizing over $\theta$ and $\delta$:
\begin{equation}
\label{eqn:calpostpred}
p_{\rm Emulate}(f_k(\tilde{x})\vert {\bf C}_k,{\bf D})
\equiv \int_{\theta,\delta} p_{\rm Emulate}(f_k(\tilde{x})\vert {\bf C}_k,{\bf D},\theta,\delta)p_{\rm Calibrate}(\theta,\delta\vert {\bf C}_k,{\bf D})d\theta d\delta.
\end{equation}
As before, the first integrand shown in the posterior predictive distribution encodes how the probabilistic extrapolation is performed. However, unlike in pure emulation, this extrapolation now additionally depends on the estimated $\theta$ and $\delta.$

A {\em calibrated emulator} can then be used to compute the mean of $f_k(\tilde{x})$ from this posterior predictive distribution:
\begin{align}
\label{eqn:calpred:marginal}
& E[f_k(\tilde{x})\vert {\bf C}_k,{\bf D}]\\ \nonumber
&=\int_{\theta,\delta}\int_{f_k(\tilde{x})}f_k(\tilde{x})p_{\rm Emulate}(f_k(\tilde{x})\vert {\bf C}_k,{\bf D},\theta,\delta)p_{\rm Calibrate}(\theta,\delta\vert {\bf C}_k,{\bf D})df_k(\tilde{x})d\theta d\delta.
\end{align}
This mean is marginalized over $\theta$. We can, of course, also use the posterior predictive distribution to compute the mean of $f_k(\tilde{x})$  for a specific value of $\theta$:
\begin{equation}
\label{eqn:calpred:giventheta}
E[f_k(\tilde{x})\vert {\bf C}_k,{\bf D},\theta]
=\int_{\delta}\int_{f_k(\tilde{x})} \! \! f_k(\tilde{x})p_{\rm Emulate}(f_k(\tilde{x})\vert {\bf C}_k,{\bf D},\theta,\delta)p_{\rm Calibrate}(\theta,\delta\vert {\bf C}_k,{\bf D})df_k(\tilde{x}) d\delta.
\end{equation}

In the ideal case that $\delta=0$ (i.e., there is no unaccounted-for physics) and we can observe the real-world process without measurement error ($\varepsilon=0)$, then in the GP setting with a mean-zero assumption \cite{KoH}, the mean of the predictive distribution (\ref{eqn:calpred:giventheta}) takes the form of  a linear combination of the observations and model evaluations
\begin{eqnarray}
\label{eqn:calpred}
  E[f_k(\tilde{x})\vert {\bf C}_k,{\bf D},\theta,\delta=0]
  =\sum_{i=1}^nw_i^f(\tilde{x},\theta)y_i
    +\sum_{i=1}^m w_i^c(\tilde{x},\theta)f_{ki} ,
    \end{eqnarray}
in which the (unnormalized) weights $w$ depend on the cross-covariances between real-world observations and physics model outputs via the calibration parameter(s) $\theta$ and input $\tilde{x}.$  The calibrated predictions therefore inherit useful information from the model outputs {\em if} the calibration parameter is well estimated and the simulator outputs are not ``too far'' from the real-world observables.  But if those two conditions are not met then the second set of weights become small ($w_i^c(\tilde{x},\theta)\rightarrow 0)$ and the predictions increasingly behave as if one were simply regressing on the observations ${\bf y}$, i.e., they ignore the physics-model outputs ${\bf f}_k$.  Note that this behavior is analogous to the motivating example described in Sec.~\ref{subsec:stat_example}, and in particular Eq.~\eqref{eqn:fstarexample34}.


The priors $p(\theta)$ and $p(\delta)$ are critically important elements to understand in calibration models \cite{KoH,brynjarsdottir2014learning}.  The former encodes our information about the calibration parameter vector before we observe our observables, while the latter encodes any information we might have on unaccounted physics in our physics model. 
Though there are some identifiability concerns when including $\delta$ in our statistical model \cite{tuo2015efficient}, the challenges appear surmountable with careful modeling practices \cite{plumlee2017bayesian,plumlee2019computer}.


\begin{figure}[thp!]
\centering
\includegraphics[scale=0.19]{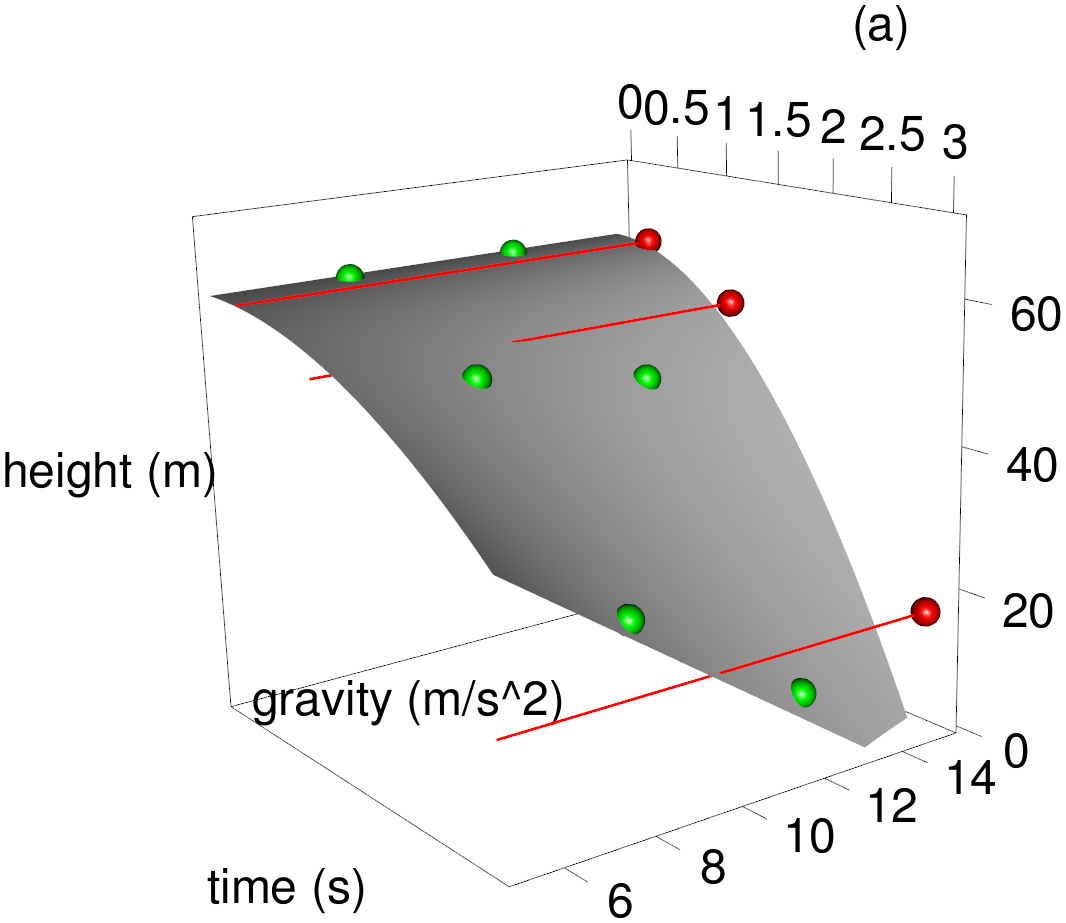}
\includegraphics[scale=0.19]{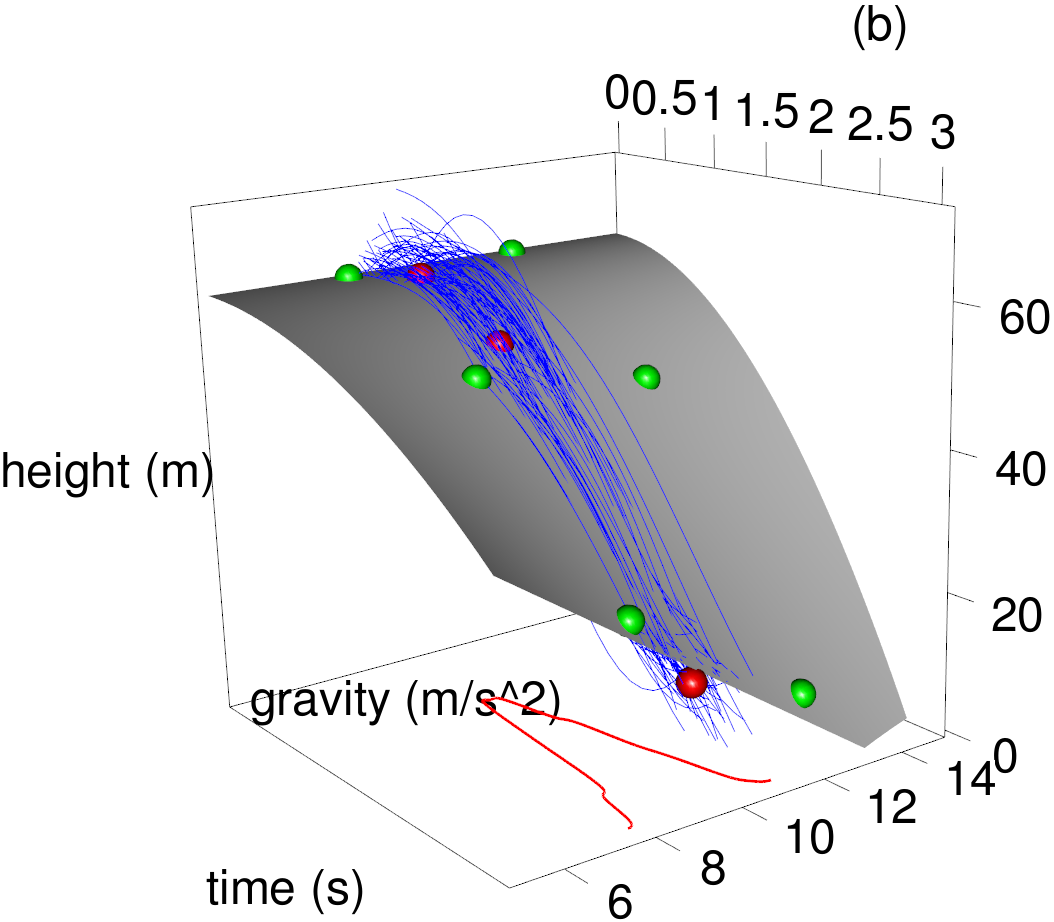}
\includegraphics[scale=0.30]{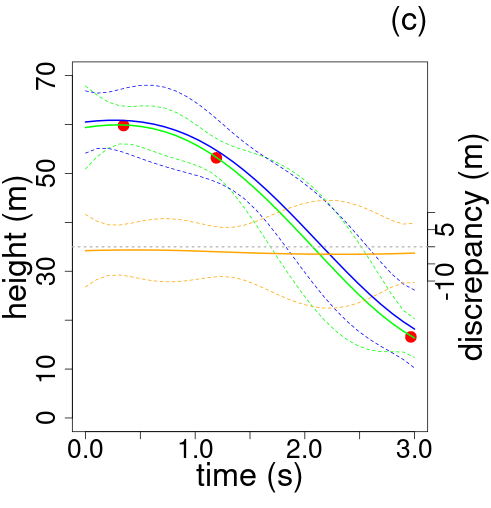}
\caption{(a) Emulation of model space.  The gray surface represents what the physics model in ${\mathcal M}_1$ would be were it available in closed-form.  The green dots represent our actual information about $f_1$, as simulated on computer.  The red dots represent the observed field observations of the real-world process. (b) The blue lines represent posterior samples of the calibrated emulator, which combines sparse information about model space $\mathcal{M}_1$ with sparse field observations to estimate the drop trajectory of the calibrated emulator for $\mathcal{M}_1$.  The red density in the z=0 plane represents the posterior estimate of the calibration parameter, gravity. (c) The corresponding calibrated emulator and its uncertainty is denoted by the blue lines.  The orange line denotes the estimated discrepancy between the model and reality, which contains 0 in its uncertainty interval across the range of time.  The corresponding predicted trajectory that combines the calibrated emulator and discrepancy is shown in green.}
\label{fig:toy}
\end{figure}

The idea of calibration is depicted graphically in Fig.~\ref{fig:toy}, where we have demonstrated the technique using the GP models for $p_\mathcal{\rm Calibrate}$.   In panel (a), the grey surface represents what the physics-model response would be in $\mathcal{M}_1$.  In practice, we only sparsely compute  $f_1(x_i,\theta_i)$ at a finite collection of input settings $\lbrace x_i,\theta_i\rbrace_{i=1}^{m_1}$ as denoted by the green dots. These form our vector ${\bf f}_1.$  The observables ${\bf y}$ are displayed as red dots (here simulated from $\mathcal{M}_1$ at $g=9.8$ m/s$^2$), however in the context of the model space of $\mathcal{M}_1$ we do not know where the red dots are located since $\theta(=g)$ is unknown. Hence the red dots should really be thought of as the red lines (i.e., the observations could correspond to any value of $\theta$ a priori).  Panel (b) displays the inferences made using calibrated emulation of $\mathcal{M}_1$. The red curve in the $x{-}y$ plane denotes the posterior density of $\theta$ and the blue lines are realizations of the posterior predictive distribution.  Note that the spread of the blue lines conveys the impact of the multiple sources of uncertainty on our inference: the uncertainty in $\theta$ as well as the uncertainty in the noisy observations ${\bf y}$ and the incomplete (sparse) information about $\mathcal{M}_1$ provided by ${\bf C}_1.$  Panel (c) projects this information back down to the $x{-}f_1$ plane, which is the view one would usually plot.  Here, the calibrated model's posterior mean is shown as the green line, while the mean of the inferred discrepancy  is denoted by the orange line.  The mean of the calibrated predictor is again shown in blue.

\subsection{Model mixing}

Bayesian solutions to statistical modeling problems typically involve some type of weighted average.
For instance, the Bayesian solutions to emulation and calibration described so far, e.g.,  Eqs.~\eqref{eqn:calpred},\eqref{eqn:calpred:marginal}, all share a common form: the posterior distribution of interest, e.g.,  Eq.~\eqref{eqn:calpostpred},  can always be expressed as a combination of our prior knowledge  weighted by the data-based evidence encoded in the likelihood.  The BMA outlined in Sec.~\ref{subsec:BMA}  also involves a combination, it's just that Eq.~\eqref{eqn:posteriorBMA} describes a finite linear combination rather than the continuous version seen in Eq.~\eqref{eqn:calpostpred} for the calibration model.  

The multi-model setting raises tricky questions about how, or whether, we want to average---questions we do not encounter within fixed-model statistical inference.
For example, in the simple ball-drop example, the BMA approach to the problem fits each model separately before averaging the two of them.  But the parameter $g$ is common between both models and has the same interpretation in each.  This raises several questions, for instance: might estimates of $g$ benefit from a joint approach to modeling $\mathcal{M}_1$ and $\mathcal{M}_2$?  And how do separate estimates of such models affect uncertainty quantification in comparison to joint approaches?  As mentioned earlier, BMA is optimal in the $\mathcal{M}$-closed setting, but in our $\mathcal{M}$-open reality, and particularly in a data-poor context, we may benefit from considering models jointly. 

Beyond the flexible software architecture to be developed in the BAND project, a core area of methodological research for BAND will be to explore such complexities that arise in the multi-model setting.  
For now, we outline two different solutions to our multi-model ball-drop problem, one that uses BMA and one employing a Bayesian calibration setup. This allows us to highlight some of the differences.

\subsubsection{Model mixing via BMA}

\begin{figure}[ht!]
\centering
\includegraphics[scale=0.3]{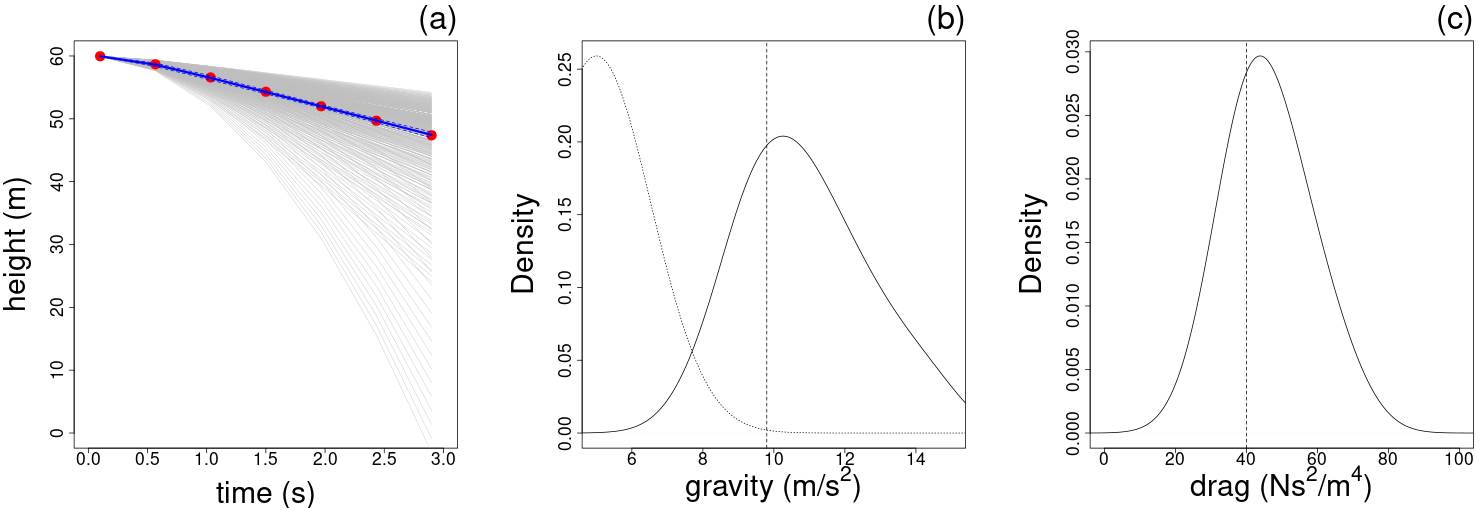}
\caption{(a) Realizations of the 2-parameter quadratic drag model in $\mathcal{M}_2$ over a $20\times 20$ grid of gravity ($g$) and drag ($\gamma$). The gray lines represent the height trajectory as a function of time for this $20\times 20$ grid of parameter settings.  The $n=7$ observations are shown as red dots while the BMA prediction and corresponding 95\% credible interval are shown in blue.  (b) The corresponding BMA density estimate for gravity, $\theta_1=g,$ and (c) the BMA estimate of the drag coefficient, $\theta_2=\gamma.$  The true values of the parameters for this simulated data are shown as the vertical dotted lines in (b) and (c).  The BMA density estimate for gravity using the wrong model ($\mathcal{M}_1$) is shown as the dotted line in panel (b).}
\label{fig:toyBMA}
\end{figure}

In a data-rich setting where the physics simulator of the real-world process can be cheaply sampled at the same  inputs as the observational data, emulation may not be needed. The BMA approach outlined in Sec.~\ref{subsec:BMA} can then be applied directly.  In this case, we have our $K=2$ models $\mathcal{M}_1,\mathcal{M}_2$ where $\mathcal{M}_1$ is equivalent to $\theta=(g,0)$ and $\mathcal{M}_2$ is equivalent to $\theta=(g,\gamma).$ The  observations are then modeled by each of these in turn, 
and we approximate the BMA solution described in Eq.~\eqref{eqn:posteriorBMA} by 
performing the model average over a discretization of $\theta$-space (alternatively, the MCMC algorithm of \cite{BMA} could be applied were $\theta$ of higher dimension).  Note that the weights for $\mathcal{M}_1$ in the BMA approach do not make use of information from the $\gamma\neq 0$ outputs from $\mathcal{M}_2.$  The resulting BMA prediction and recovered estimates of the gravity and drag parameters are shown in Fig.~\ref{fig:toyBMA}.  Since we include both drag-free and draggy models in this BMA, we expect BMA to perform well.  However, to get a sense of what can go wrong we also performed BMA ignoring the draggy model which resulted in the highly biased estimate of gravity shown as the dotted density curve in Fig.~\ref{fig:toyBMA}(b).

\subsubsection{Model mixing via calibration}


By again considering the models to be continuously indexed by $\theta=(g,\gamma)$ where $\gamma=0$ is equivalent to $\mathcal{M}_1,$ it is straightforward to cast the situation of multiple models within the calibration framework.
The calibrated predictor in (\ref{eqn:calpred}) then bears a striking resemblance to the BMA form,
\begin{eqnarray}
\label{eqn:calpred:multi}
E[f_1(\tilde{x})\vert {\bf C}_1,{\bf C}_2,{\bf D},\theta,\delta=0] &=&\sum_{i=1}^nw_i^f(\tilde{x},\theta)y_i+\sum_{i=1}^{m_1}w_{1i}^c(\tilde{x},\theta)f_{1i}\nonumber\\
& &+\sum_{i=1}^{m_2}w_{2i}^c(\tilde{x},\theta)f_{2i},\end{eqnarray}
where we see that the (unnormalized) weights for the outputs of both models in Eq.~\eqref{eqn:calpred:multi} in fact depend on the parameter $\theta$ spanning both model spaces {\em and} the input setting $\tilde{x}.$  This expectation would then be further re-weighted as in Eq.~\eqref{eqn:calpred:marginal} where $p_\mathcal{\rm Calibrate}(\theta,\delta\vert {\bf C}_1,{\bf C}_2,{\bf D})$ now involves the joint posterior.  In other words, the calibration solution outlined considers both models jointly, and we can think of $E[f_1(\tilde{x})\vert {\bf C}_1,{\bf C}_2,{\bf D}]$ as approximating $E[f_1(\tilde{x})\vert \mathcal{M}_1,\mathcal{M}_2]$ and similarly $p_\mathcal{\rm Calibrate}(\theta,\delta\vert {\bf C}_1,{\bf C}_2,{\bf D})$ as approximating $p_\mathcal{\rm Calibrate}(\theta,\delta\vert \mathcal{M}_1,\mathcal{M}_2).$


\begin{figure}[ht!]
\centering
\includegraphics[scale=0.18]{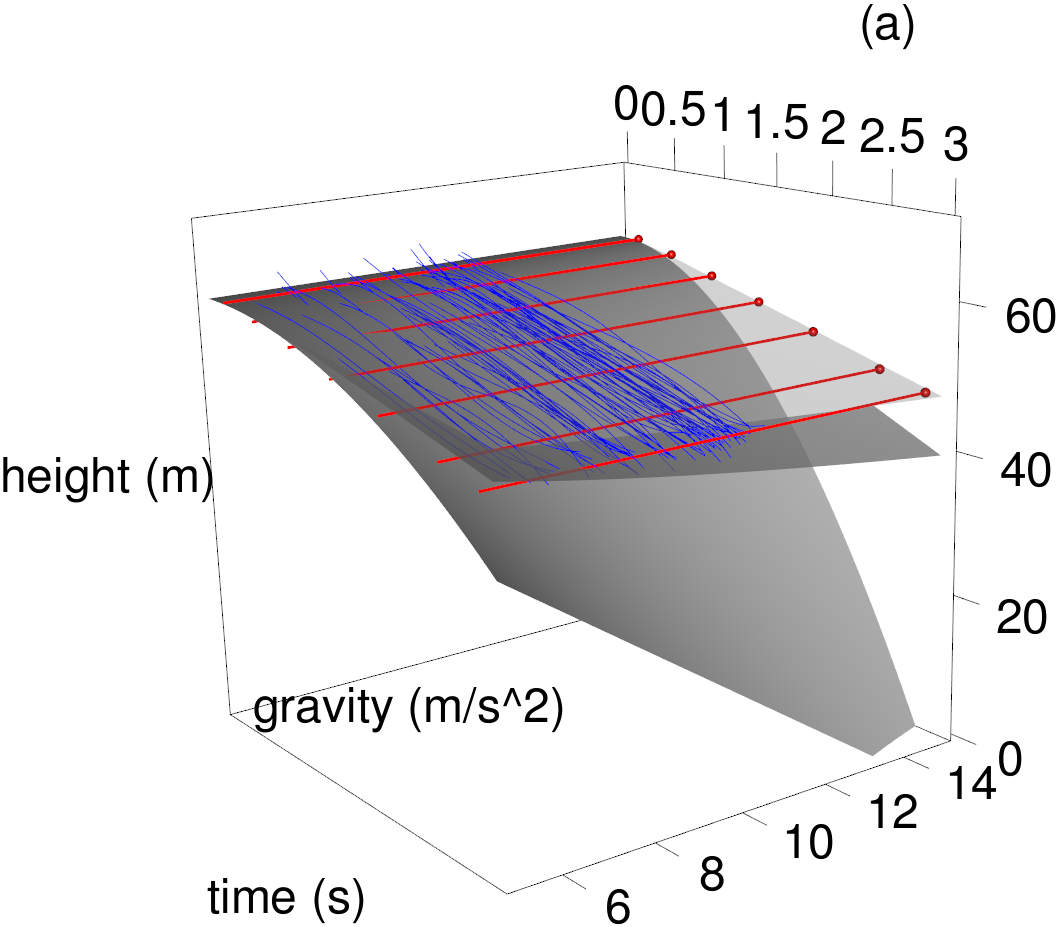}
\includegraphics[scale=0.29]{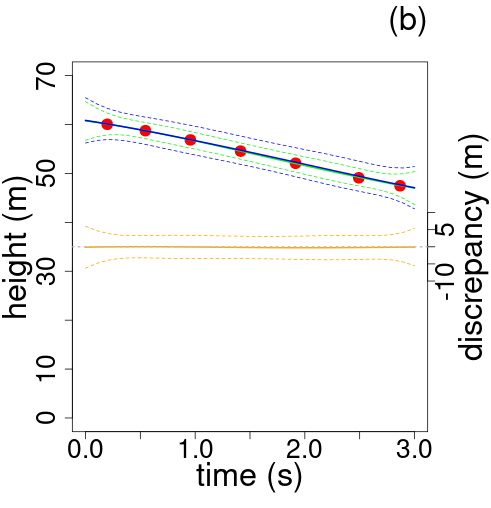}
\includegraphics[scale=0.29]{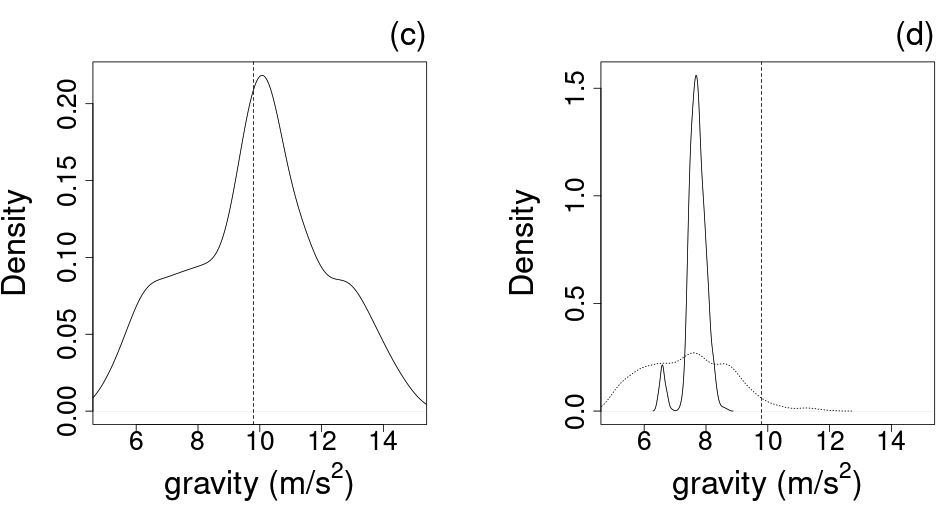}
\caption{(a) Calibration with two models: the drag-free physics model (as in Fig.~\ref{fig:toy}) and a quadratic-drag physics model.  The gray surfaces represent what the physics model would be were it available in closed form with no drag ($\mathcal{M}_1$), and with quadratic drag  ($\mathcal{M}_2$).  The quadratic drag surfaces are plotted using drag coefficients of $\gamma=25,75$.  Note the much more linear appearance of $f_2$ at these two settings of $\gamma,$ and the corresponding reduction in drop distance as compared to the drag-free model.  The model-mixed calibration is shown as the blue curves.  (b) The corresponding model-mixed calibrated emulator and its uncertainty is denoted by the blue lines.  The orange line denotes the estimated discrepancy between the model and reality, which contains 0 in its uncertainty interval across the range of time.  The corresponding predicted trajectory that combines the model-mixed calibrated emulator and discrepancy is shown in green.  (c) Posterior density of gravity ($\theta_1=g$) is shown in this multi-model setup.  (d) Corresponding posterior density of gravity when using the incorrect model $\mathcal{M}_1$ is shown here with the same discrepancy prior as in the multi-model calibration (solid line) and with a more vague prior (dotted line).}
\label{fig:toymix}
\end{figure}

A demonstration of this idea is shown in Fig.~\ref{fig:toymix}, where we now consider both our drag-free model $\mathcal{M}_1$ and the quadratic-drag model $\mathcal{M}_2$ that depends on the additional drag coefficient parameter, $\gamma$.  Setting $\gamma=0$ recovers the drag-free model, and the gray surfaces depict the physics model evaluated at $\gamma=0$ (i.e., as in  $\mathcal{M}_1$), $\gamma=25$ and $\gamma=75$ in the figure.  Note that the behavior of both models is similar up to about $x=1$ seconds, indicating that $f_1$ can still be leveraged for prediction in this regime.  However, beyond $x=1$ seconds, the models diverge significantly, indicating that information can only usefully be borrowed from $f_2$, even though $\mathcal{M}_2$ is more sparsely sampled.  The observations were generated with a drag coefficient of $\gamma=40$ at $n=7$ time points, as denoted by the red dots in Fig.~\ref{fig:toymix}(b).  We see that even though $\mathcal{M}_1$ is not meaningful beyond $x=1$ seconds and $\mathcal{M}_2$ is much more sparsely sampled than the drag-free model, the overall prediction is well behaved. 
The resulting posterior for gravity ($g$) shown in Fig.~\ref{fig:toymix}(c) is well centered on the true value.  Meanwhile, calibrating only using $\mathcal{M}_1$ (the {\em incorrect} model) results in the biased estimates shown in Fig.~\ref{fig:toymix}(d) for both strong and weak priors on the discrepancy.

\subsection{Experimental design questions}

Within the toy model we can imagine a range of enhanced experiments to better measure the gravitational constant $\theta$: build a taller tower to reach greater ball speeds (``energy frontier'') or develop better clocks and rulers (``precision frontier'') or drop more  balls (``intensity frontier'').
Deciding which option to pursue and with what specifications is a problem of experimental design.
We turn to the Bayesian approach to this problem in the next section.


\section{Experimental design} 
\label{sec:design}

Bayesian experimental design provides a framework in which experiments can be designed using the current information available both from experiment and theory.
Broadly speaking, NP experiments involve a plethora of observables measured with a  great variety of techniques, ranging from simple decay and scattering experiments to cross-section reactions with radioactive ion beams, to relativistic heavy-ion collisions. Experiments can be expensive, and communities often have to choose between competing proposals for new apparatus or for beam time. 

To optimize experiments, the goals of the experimenter are encoded in a \emph{utility function} which describes the usefulness of potential observations and may also include the cost of the experiment. One then considers various future \emph{experimental designs} and computes the \emph{expected utility} of each design by averaging over all potential experimental results from that design. 
A particular \emph{experimental design} might be specified by an observable and a set of experimental conditions at which to measure it (e.g., beam energies and detector positions) and perhaps also the experimental noise levels. 
Experimental  regimes (e.g., kinematic regions) where limitations of the  facility being used for the experiment are liable to make collecting data excessively difficult can be excluded from the optimization by explicit restrictions on the designs considered. Once the utility function and the possible designs have been specified, the optimal design is simply the scenario that  maximizes the expected utility function over the domain of possible designs. 

In order to invoke the experimental design formalism, the goal of the experiment must be specified. Is it to make an accurate observation of some quantity? To discriminate between competing models? Or to precisely constrain parameters of the theory? In this section we illustrate the Bayesian approach to experimental design by focusing on experiments with the last of these three goals. We define the optimal design as the one which provides the greatest increase, on average, in the knowledge of the parameters of the NP model. The state of knowledge about those parameters before any new experiment is performed is incorporated in our experimental design  using Bayesian priors.

In general the experimental  goal is encoded as a utility function, or design criterion, $U(\design, \qoi , \genobsset)$, that depends on the design points\footnote{%
    A single design (observable, experimental conditions, etc.) is denoted by $\design$. The space $E$ is the set of all considered experiments over which the utility is optimized (\eg, all possible 5-angle measurements of a differential cross section at a given energy).}
$\design$ in the design space $E$ from which experimental data $\genobsset$ are then measured and the quantities-of-interest $\qoi$ that we have constructed our experiment to find. Of course, $\genobsset$ will not be known until the experiment is conducted. Hence the optimal design $\design^\star$ is that which maximizes the \emph{expected} utility $U(\design) = E[U(\design, \qoi, \genobsset)]$. In this section we focus on the case where $\qoi$ are the (physics-model) parameters $\theta$, so we seek
\begin{align}
    \design^\star & = \argmax_{\design \in E} U(\design) \notag \\
    & = \argmax_{\design \in E} \int U(\design, \theta, \genobsset) \pr(\theta, \genobsset \given \design)\, \dd{\theta}\, \dd{\genobsset} \label{eq:expected_utility} \\
    & = \argmax_{\design \in E} \int \Big\{ U(\design, \theta, \genobsset) \pr(\theta \given \genobsset, \design)\, \dd{\theta} \Big\} \pr(\genobsset \given \design)\, \dd{\genobsset}  \notag
\end{align}
where $\displaystyle{\argmax_{\design \in E}}\  U(\design)$ denotes the maximum of the utility function over all choices of ${\design \in E}$. For each possible experimental outcome $\genobsset$, we compute corresponding posteriors for the parameters $\theta$. By then marginalizing over $\genobsset$, with a weighting given by the probability of that $\genobsset$ for a given $\design$ as predicted by the model or its emulator, we average the expected gain in information on the parameters $\theta$ over all data that could plausibly be measured. To sample all those possibilities is often computationally quite expensive, which is why emulators are a key part of the BAND framework. However, if the predictions can be reliably linearized around the best known parameters then a simple and intuitive formula for the expected utility of an experiment is obtained \cite{Melendez:2020ikd}.

Equation~\eqref{eq:expected_utility} says that the process of experimental design requires a theory $f(\kinparvec, \theta)$ and a probabilistic model relating data to theory parameters, $\pr(\theta, \genobsset \given \design)$. To calculate that pdf we use the product rule to  write $\pr(\theta, \genobsset \given \design) = \pr(\genobsset \given \theta, \design) \pr(\theta)$ (likelihood for given design $\times$ prior). To evaluate the likelihood $\pr(\genobsset \given \theta, \design)$ we need to include the theoretical model discrepancy in a model such as Eq.~(\ref{eq:stand_model}). Here we'll use for illustration a Gaussian prior and (correlated) Gaussian errors in the model (e.g., see Ref.~\cite{Melendez:2020ikd}). We suppose that at the start of our experimental-design process prior knowledge of the parameters of interest is specified by a multi-variate normal distribution with a vector of means $\mu_0$ and a covariance matrix $V_0$,
\begin{align} \label{eq:polarizability_prior}
    \pr(\theta) = \normal(\mu_0, V_0) \, ,
\end{align}

Under the assumption that $f(\kinparvec,\theta)$ is linear in $\theta$, it follows that the posterior is also given by a normal distribution
\begin{align} \label{eq:polarizability_posterior}
    \pr(\theta \given \genobsset, \design) = \normal(\mu(\genobsset, \design) ,  V(\design)) \, ,
\end{align}
where the mean and variance have been updated from $\mu_0$ and $V_0$ to $\mu(\genobsset, \design)$ and $V(\design)$ respectively. Crucially, $V(\design)$ depends on neither the
specific value of $\mu_0$ nor the measured data $\genobsset$. Instead the extent to which it updates $V_0$ is determined by a combination of 
the model error and the experimental errors.


The optimal design is then that which provides the best improvement in constraints on $\theta$, i.e., the greatest improvement in $V$ over $V_0$. This leads us to choose the utility to be the gain in Shannon information compared to prior information for $\theta$, based on the experiment $(\design, \genobsset)$. This is equivalent to the so-called Kullback-Leibler (KL) divergence, or relative entropy, between the prior and posterior for $\theta$ (a measure of the difference between these probability distributions), followed by marginalizing over $\genobsset$:
\begin{align}
    U_{\text{KL}}(\design)
    & = \int \!\!\left\{\! \ln\!\!\left[\frac{\pr(\theta \given \genobsset, \design)}{\pr(\theta)}\right]\! \pr(\theta \given \genobsset, \design)  \dd{\theta}\right\}\! \pr(\genobsset \given \design) \dd{\genobsset}\!. \label{eq:expected_utility_kl}
\end{align}

In fact, if linearization is valid, the integral over $\genobsset$ is trivial since neither the posterior not the prior covariance matrix depend on it. Equation 
 \eqref{eq:expected_utility_kl} can be computed exactly
(see Appendix~A of Ref.~\cite{Melendez:2020ikd}),
with the result
\begin{align} \label{eq:utility_kl_analytic}
    U_{\text{KL}}(\design) = \frac{1}{2} \ln \frac{|V_0|}{|V(\design)|} \equiv \ln\shrinkage(\design) \geq 0 \, ,
\end{align}
where we have defined the posterior shrinkage factor $\shrinkage \geq 1$. Our assumptions lead to a form of the expected utility that is analytic, easy to understand, and quick to compute. Particular confidence levels for the prior \eqref{eq:polarizability_prior} and posterior \eqref{eq:polarizability_posterior} for the parameters $\theta$ define hyperellipsoids. Then $\shrinkage$ is the factor by which the volume of the prior ellipsoid shrinks as it is updated to the posterior, with larger values of $\shrinkage$ (or $U_{KL}$) being more informative than smaller values.
An experiment yielding $\shrinkage = 1$ (or $U_{KL} = 0$) is then completely uninformative.
If we are interested only in a subset of the $\theta$, and not in the rest, we can re-define the utility to find the optimal design of an experiment that seeks to measure our subset of interest by simply computing  Eq.~\eqref{eq:utility_kl_analytic} with the corresponding submatrices of $V_0$ and $V$.

\begin{figure}[!tbp]
    \centering
    \includegraphics[width=\textwidth]{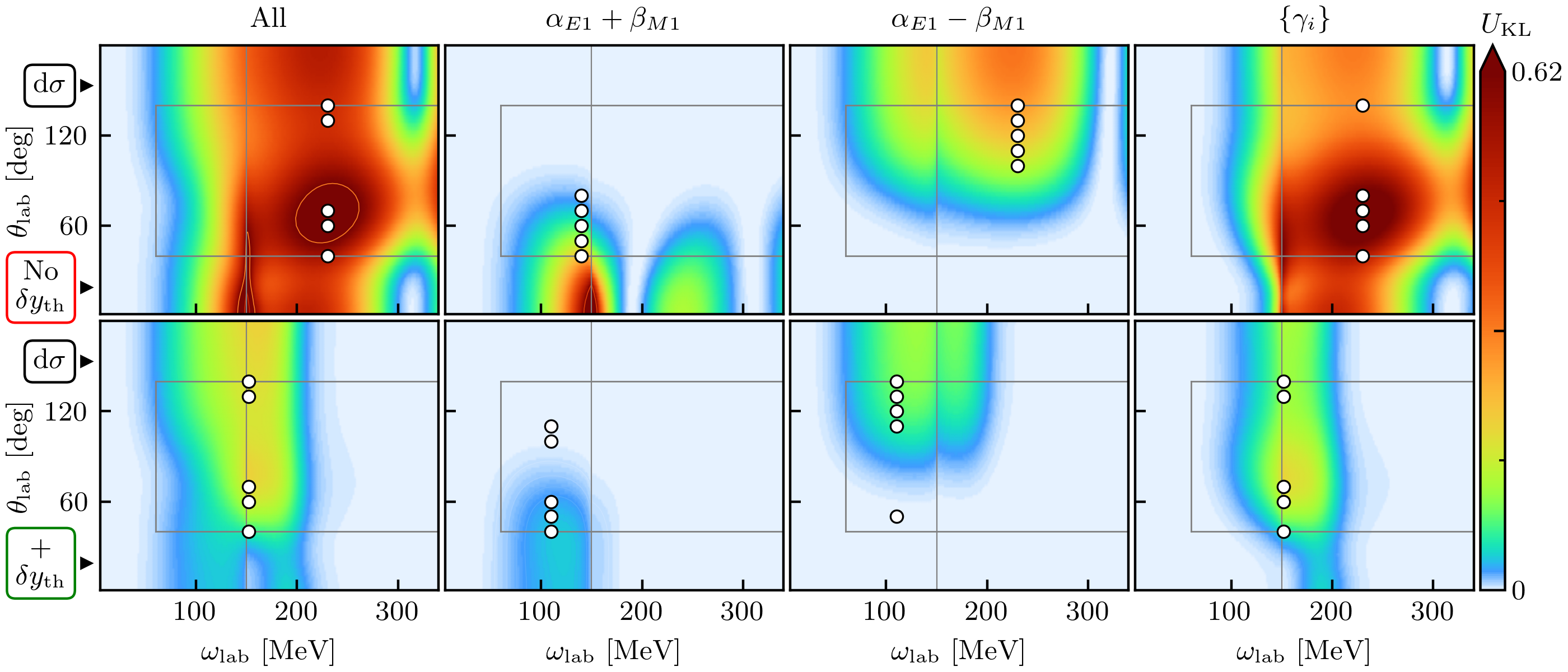}
    \label{fig:utility_grid_compare_subsets_and_truncation_proton}
    \caption{Example illustrating the concept of experimental design. The expected utility  Eq.~$\eqref{eq:utility_kl_analytic}$ of proton differential cross section ($\diffcs$) measurements (see Ref.~\cite{Melendez:2020ikd} for details).
    Colors indicate the utility of one measurement conducted at each kinematic point $(\omegalab, \thetalab)$, with the point of largest utility $U_{KL}$ being by definition the optimal 1-point design. (The color bar is on a linear scale, though the hue varies \emph{much} more quickly for small $U_{KL}$.)
    The top row (with the red, ``No $\delta\genobsth$'' box) does not include model truncation estimates, whereas the bottom row does include this uncertainty.
    Each column shows the information gain one could expect to achieve for a subset of the proton 
    polarizabilities. 
    The white circles with black borders show the optimal design kinematics for five measurement points at the same energy but different angles. Reproduced from Ref.~\cite{Melendez:2020ikd} with kind permission of The European Physical Journal (EPJ).
    } 
    \label{fig:utility_grid_compare_subsets_and_truncation}
\end{figure}


Note that constraints from previous experiments are built in naturally via the prior on the parameters. So, if we find a large utility in an observable or a region of experimental conditions that has already been thoroughly explored, that means there is still valuable constraining information to be gained there.

As an illustrative example, Fig.~\ref{fig:utility_grid_compare_subsets_and_truncation} shows the expected utility from Eq.~(\ref{eq:utility_kl_analytic}) for experiments to measure Compton scattering from the proton \cite{Melendez:2020ikd}. Each panel in the top or bottom row shows a color contour plot of $U_{\text{KL}}(\design)$ at possible kinematic points (specified by laboratory energy and scattering angle) for determining a subset of proton polarizabilities from the measurement of the proton differential cross section (see Ref.~\cite{Melendez:2020ikd} for further examples and explanations). The polarizabilities are extracted through application of a NP model (here: chiral effective field theory). The most red regions are where the most fruitful measurement will be. The top row does not include the theoretical model discrepancy, which in this case is from the model truncation error, while the bottom row does include this uncertainty. The effect of including the truncation errors is striking: it shifts the region of optimal utility to lower energies and moderates the expected information gain. Including theory uncertainties is essential for experimental design!

Now suppose we have multiple models. Then our observational conditional, $p(\genobsset|\design,\theta)$, will be replaced by a mixed model conditioning. For example, if BMA is used for the mixing then the mixed model for observables can be formulated according to Eq.~(\ref{eqn:posteriorBMA}). As long as the parameters $\theta$ are common to all models used in the mixing we can employ the above formalism by revising $\pr(\theta, \genobsset \given \design)$ accordingly in Eq.~(\ref{eq:expected_utility}).  However, the use of general mixing can lead to more complicated forms than the illustration presented here. Such use of model mixing for experimental design is one of the ultimate goals of the BAND project.  
 
\section{Case Study: The equation of state of strongly interacting matter} \label{sec:eos}

Heavy-ion collisions, performed at energies from a few MeV to a few tens of TeV provide the means to excite femtoscopic regions of  matter to extreme densities and temperatures. Great experimental investments have been made at NSCL \cite{NSCL}, RIKEN \cite{RIKENNISHINA}, GSI \cite{GSI},   RHIC \cite{RHIC}, and LHC \cite{LHC} to explore strongly interacting matter at temperatures from a few to hundreds of MeV and densities up to several times nuclear matter density. New facilities are coming online, as FRIB \cite{aFRIB}, FAIR \cite{FAIRCBM}, and  NICA \cite{NICA} should all be completed in the next few years.

Although these experiments address a wide variety of issues, two critical areas of commonality will be addressed by BAND. First, existing and future high-quality datasets are enormous and cover a remarkably heterogeneous range of physics by employing a vast complement of detectors. Secondly, the created hot and dense matter cools quickly and is very short-lived, and interpreting the measurements thus requires comparison to sophisticated and numerically intensive theoretical models and simulations describing its evolution through multiple stages before being observed. These models build on robust theoretical frameworks for describing strongly interacting matter in its various manifestations but involve a number of parameters describing medium properties that cannot yet be precisely computed from first principles. In addition, the transitions between different stages provide conceptual challenges that result in competing models built on conflicting paradigms, assumptions and/or approximations. BAND's role lies at the intersection of experiment and theory where comprehensive experimental datasets are analyzed using Bayesian inference to constrain the uncertainties in model structure and model parameters. Given the complexity of these model-to-data comparisons, sophisticated new methodologies from the statistical science community are required to achieve complete and rigorous uncertainty quantification including both experimental and theoretical sources of error.

\begin{figure}[htb!]
\begin{center}
\includegraphics[width=0.6\textwidth]{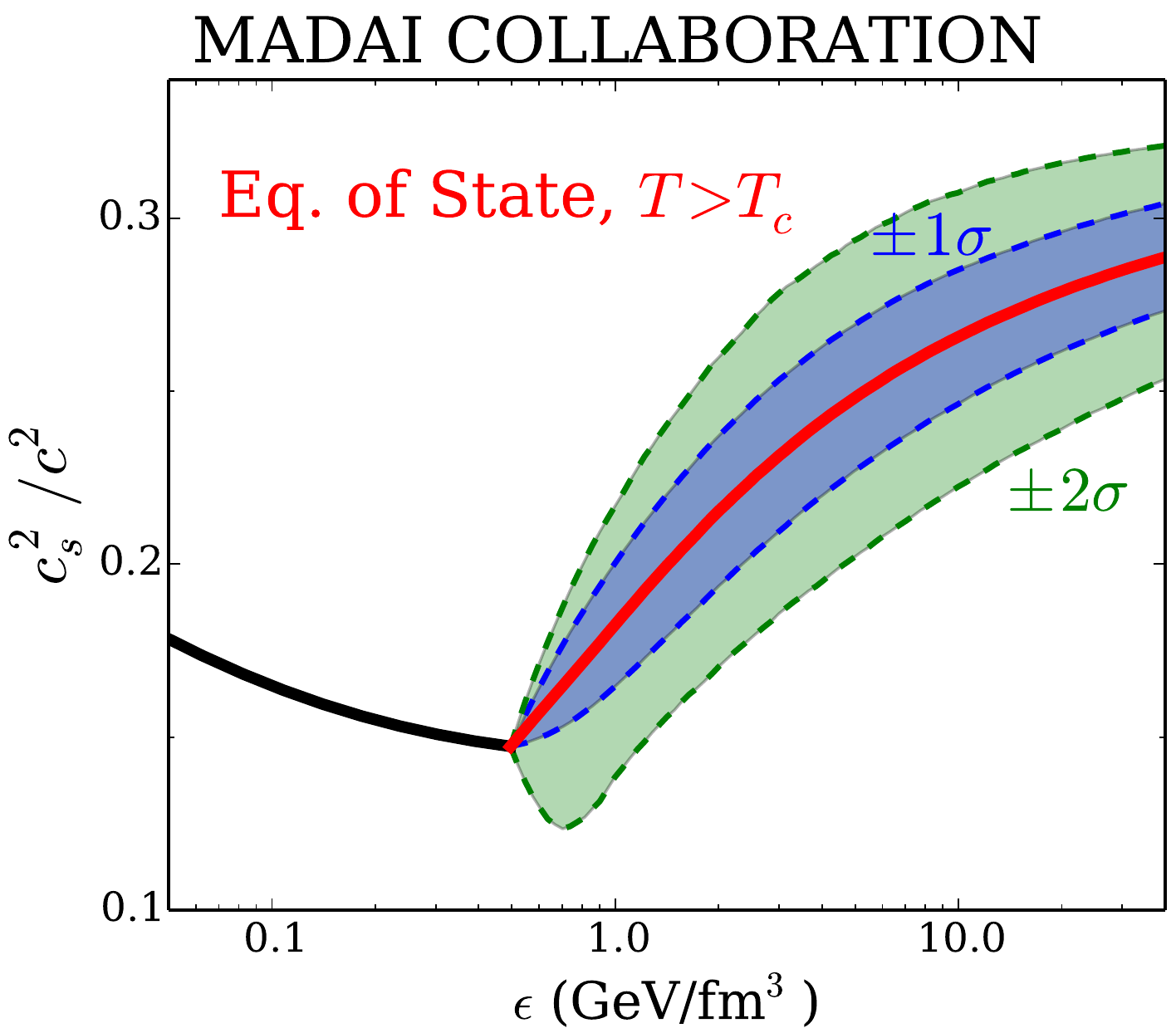}
\end{center}
\caption{The speed of sound vs. energy density for baryon-free matter as constrained by a 14-parameter model compared to data from RHIC and the LHC. Results are described in detail in Ref.~\cite{Pratt:2015zsa}.}
\label{fig:madai}
\end{figure}

Statistical approaches based on model emulators have recently been applied to analyses of heavy ion data from RHIC and the LHC \cite{Pratt:2015zsa, Paquet:2020rxl}. After being tuned using a few hundred to several thousand full model runs at each point of a sufficiently large number of design points for the model parameters, emulators reproduce principal components of the model output (predictions for observables) with little computation. This enables  exploration of the high-dimensional parameter space with fine resolution for mapping out the joint posterior distribution for the model parameters. These analyses result in likelihood contours of the parameter space where uncertainties, both experimental and theoretical, are taken into account. The result of one such analysis performed by the MADAI Collaboration \cite{madai} is presented in Fig.~\ref{fig:madai}. Here, a 14-dimensional parameter space was explored in analyzing high-energy collisions from RHIC and from the LHC \cite{RHIC,LHC}. Parameters expressing the equation of state were among those varied, and the ensuing constraint of the equation of state is shown in the figure.

Going forward, a main challenge facing the field is to handle multiple competing models that do not necessarily share a common set of parameters. All applications of emulators to heavy-ion collisions to date have accounted for parameter variation within a particular model. However, there are instances where multiple models must be simultaneously considered. For heavy-ion collisions this is especially true for models of the initial stopping stage for lower energy collisions corresponding to the RHIC Beam Energy Scan, for the pre-hydrodynamic evolution, and for the interface between the hydrodynamic and late hadronic simulation stage (for this last issue see Sec.~\ref{sec:BMA_TC}.) For the initial conditions and pre-hydrodynamic stage, several models based on very different paradigms should be considered. Both for the purpose of determining the best choice of early-stage models, and for accurately reflecting the uncertainty in the early evolution stage when extracting information about the medium properties controlling the hydrodynamic stage of the collision, one must consider a variety of theoretical pictures. This challenge defines the principal role of BAND's expertise in applications to heavy-ion physics.

\section{Case Study: Design of experiments for nuclear reactions} \label{sec:reactions}

The Facility for Rare Isotope Beams  will come online soon and will offer the possibility of producing thousands of rare isotopes, many of which are unobserved and extremely neutron rich. Due to the complexity of each experiment, the facility cannot (and should not!) measure them all. Reactions offer an array of probes into the structure of nuclei. Reactions at FRIB will also be used as indirect methods for extracting reaction rates for astrophysics \cite{nunes2020}.  In planning for these future experiments we can ask: What are the best beam energies? What is the required angular range? Which reaction products should be detected? What reaction observables should be measured? Etc. As discussed in Sec.~\ref{sec:design} the answers to these questions will depend on the goal; once a goal has been chosen it can be encoded in a utility function.

Due to the complexity of an ab-initio theory for reactions involving intermediate mass and heavy nuclei, few-body models are commonly used. In these models, most nucleonic degrees of freedom are frozen, and only a few are included in the dynamics. In such cases, the essential ingredient to the calculations becomes the optical potential: an effective complex interaction between the relevant composite bodies that captures the many-body complexity of the problem. Nucleon-nucleus optical potentials have been traditionally obtained from fitting data, primarily elastic scattering. Global optical potential parameters (e.g., \cite{bg69,kd2003}) obtained using  standard $\chi^2$ minimization \cite{lovell2017} are charge, mass and energy dependent and only provide an average description of reactions across the nuclear chart. Indeed, particularly for reactions with unstable nuclei, the accuracy of global approaches is unknown due to the extrapolations to nuclei far away from the valley of stability. To properly leverage the massive investment of time, scientific expertise, and resources we must understand how the uncertainties in  models that are fitted to data propagate to their predictions, and especially to extrapolated predictions for targets with extreme neutron or proton numbers. 

In the last few years, Bayesian methods have been established to quantify the uncertainties in the optical potential parameters and corresponding  observables \cite{lovell2018,king2019}.  Initial work in Refs.~\cite{lovell2018,king2019} focused on how well a single set of elastic scattering data characterized by a well defined beam energy and a generous angular distribution could pin down the optical-potential parameters. Mock data were generated for elastic angular distributions using the model of Ref.~\cite{kd2003} and an overall 10\% error on these synthetic observations  was assumed. These data were then used to calibrate an optical potential model of the reaction containing 9 parameters. Wide Gaussian prior distributions centered around the global parameters of \cite{bg69} were chosen as the prior for these parameters. The nine-dimensional parameter posterior was then generated from Monte Carlo sampling using the Metropolis-Hastings algorithm. These posteriors were then used to obtain the credibility intervals for the elastic scattering angular distributions and propagated to other reaction observables such as the total (reaction) cross section and the transfer angular distribution, see Eq.~(\ref{eq:Bayesprediction}). The most striking conclusion from these Bayesian studies \cite{lovell2018,king2019} was that the resulting posterior distributions for predicted observables were significantly wider than previously assumed and did not exhibit Gaussian shapes. The linear error propagation assumed in previous studies was not valid for this situation. In fact, the credibility intervals obtained when the optical potential is calibrated on elastic data of this accuracy and results propagated to a transfer reaction are too large for a useful model comparison. These early UQ studies for optical potentials suggest that the way they are presently constrained by data leads to too much uncertainty for their application in other reactions to give significant insights into the dynamics of those reactions.

\begin{figure}[htb]
\begin{center}
\includegraphics[width=0.5\textwidth]{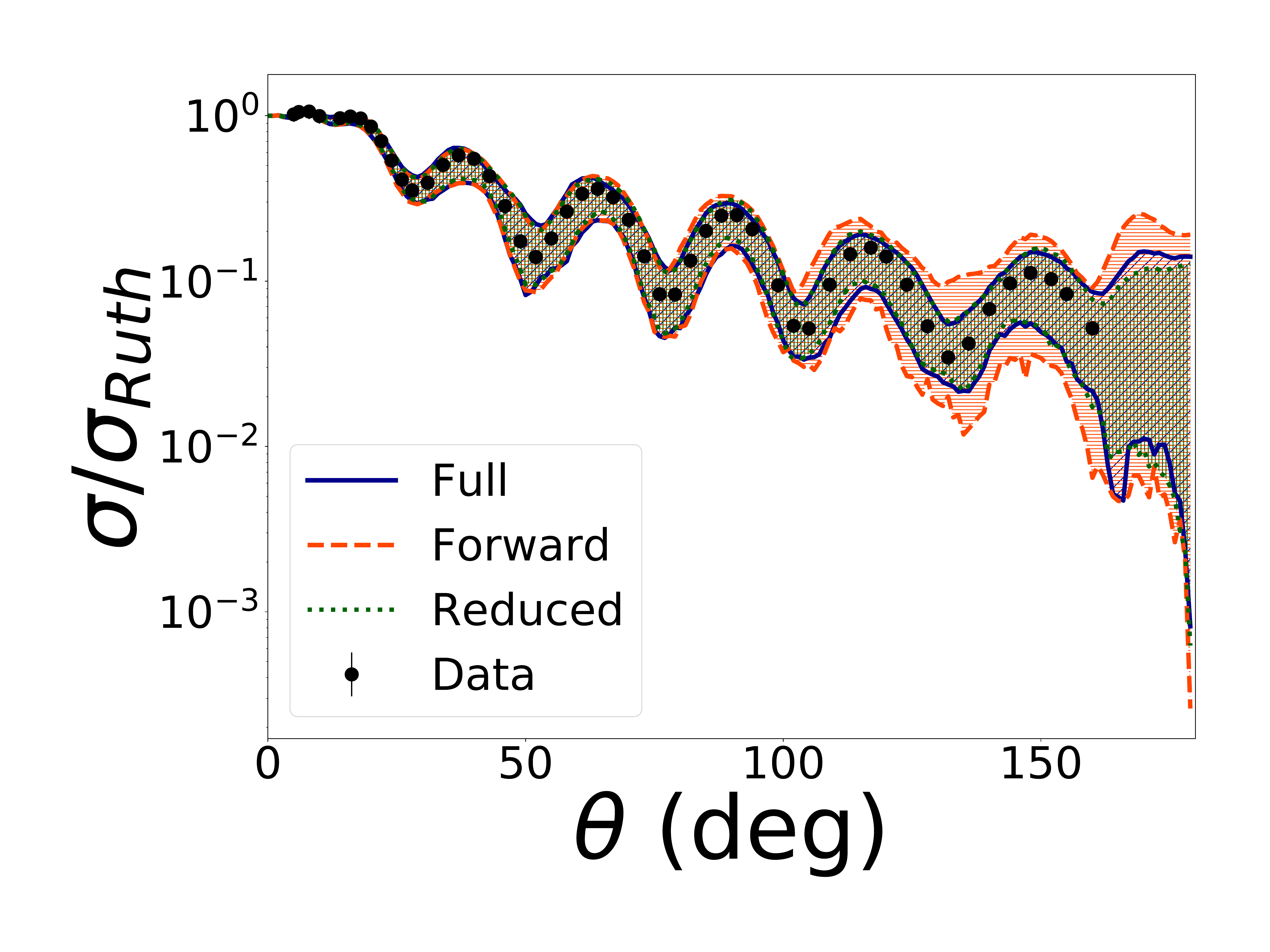}
\end{center}
\caption{Angular distributions for elastic scattering of protons on $^{208}$Pb at 30 MeV: 95\% credibility intervals when including data in the full angular range (full, blue solid line), when only including forward angles (forward, orange dashed line) and when including a sparse angular grid (reduced, green dotted line). The $y$-axis displays the ratio of the simulated proton-${}^{208}$Pb cross section to the Rutherford cross section. Results are described in detail in Ref.~\cite{catacora2019}.}
\label{fig-ppb1}
\end{figure}

\begin{figure}[htb]
\begin{center}
\includegraphics[width=0.5\textwidth]{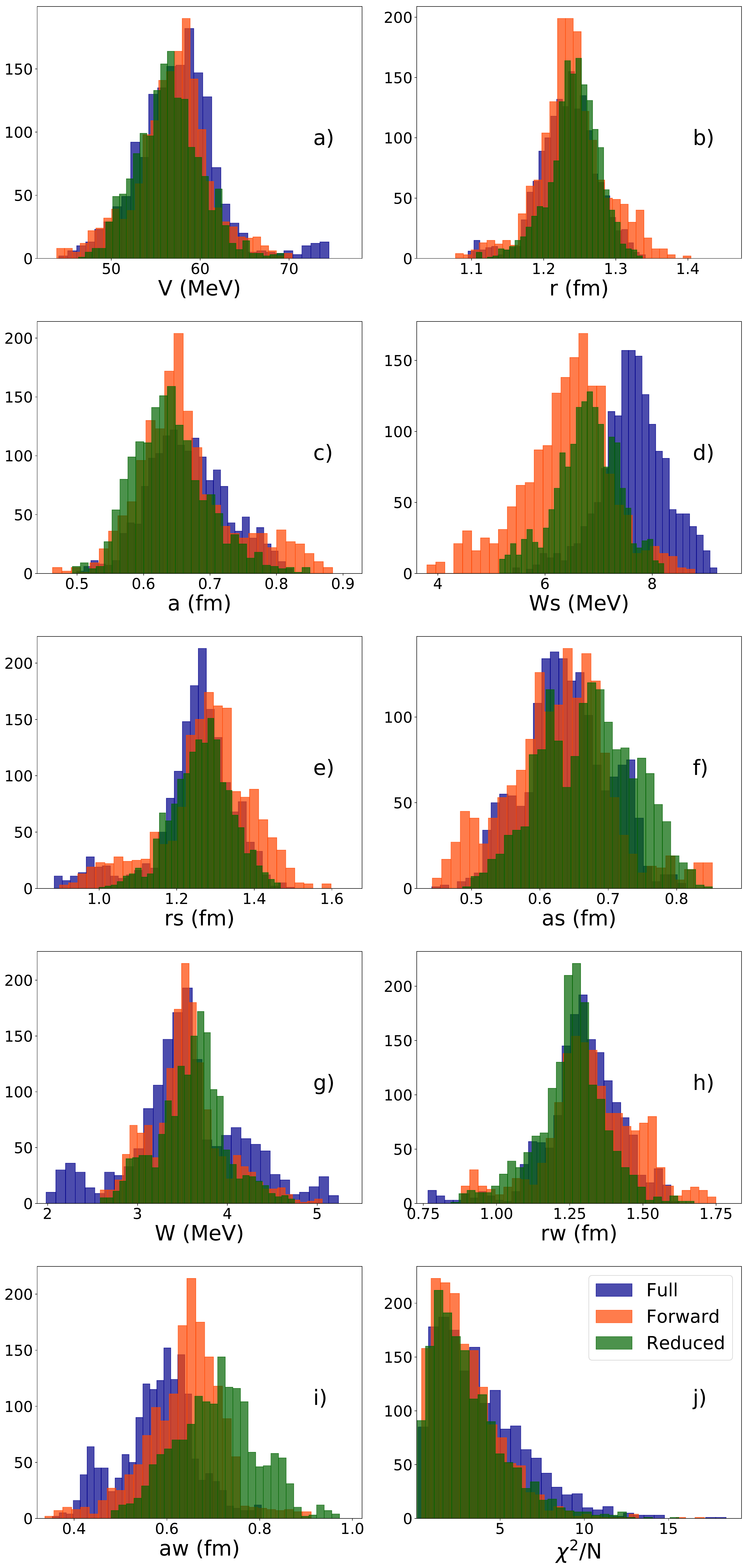}
\end{center}
\caption{Parameter posterior distributions for the elastic scattering of protons on $^{208}$Pb at 30 MeV:  including data in the full angular range (blue), when only including forward angles (orange) and when including a sparse angular grid (green)~\cite{catacora2019}. }
\label{fig-ppb2}
\end{figure}

Since optical-potential models are workhorses of nuclear-reaction theory it is important to understand how these too-large uncertainties could be reduced. Which observables and kinematic conditions can provide a significant reduction of this uncertainty? As a first step to a full experimental-design analysis Ref.~\cite{catacora2019} asked how impactful it is to reduce the experimental error. This is largely dominated by the point-to-point error for experiments with rare isotopes, so issues with discrepancy functions were not discussed in this initial study. Ref.~\cite{catacora2019} then showed that, for most cases, a factor of two reduction in the point-to-point uncertainty of observations does not result in a factor of two reduction in the uncertainty of the model prediction for the elastic angular distribution.


The angular range is also another important consideration in such experiments. As an illustration, Fig.~\ref{fig-ppb1} shows the 95\% credibility intervals obtained for the angular distributions for the elastic scattering of protons on $^{208}$Pb at 30 MeV. These are presented in terms of the ratio to elastic scattering due purely to the Coulomb interaction---the ``Rutherford cross section". This removes the divergence in the results at zero degrees.
The results obtained using the ``full'' angular distributions (180 data points from $1$ to $180$ degrees) are shown as the green band and compared to a ``reduced'' analysis when only every tenth data point is used for model calibration. The differences in the posterior predictive distribution obtained from the full and reduced dataset are imperceptible. By contrast, when only data for angles below 100 degrees is included (``forward'' analsysis) the orange band thereby obtained is markedly wider  (note the log scale) at the backward angles where constraining data were not included.

Figure~\ref{fig-ppb2} shows the corresponding posteriors for the optical-model potential parameters: the depth, radius and diffuseness of the real part of the optical potential $(V, r, a)$ and the imaginary terms, surface $(W_s,r_s,a_s)$ and volume $(W,r_w,a_w)$. The most important difference between  calibration with data over the full angular range and that which uses only forward-angle data is in $W_s$. Reference~\cite{catacora2019} concluded that using a dense angular grid in the experiment is likely a waste of resources, but there is important information in the backward angles observations that makes a substantial difference to the model calibration.

The BAND framework will be brought to bear on these issues. A first step will be to use a utility function as described in Sec.~\ref{sec:design} to quantify the notions of optimal experimental design implemented heuristically in Ref.~\cite{catacora2019}. Meanwhile, Secs.~\ref{subsec:likelihood} and \ref{sec:bmm} emphasized the importance of accounting for model imperfections in the likelihood function used for calibration. And Sec.~\ref{sec:design} and Ref.~\cite{Melendez:2020ikd} demonstrated that unless such a discrepancy function is included in the analysis the conclusions regarding the optimal experimental design may be misleading. So understanding the imperfections of different reaction-theory models and including statistical descriptions of them will be a key part of BAND's effort in this area. The reaction-theory community can also benefit from BAND's participatory approach to prior building: Sec.~\ref{subsec:priors} showed how hierarchical Bayesian models can be used to incorporate constraints, other data, and intuition on model parameters in the analysis. 

While the simplicity of the optical model made it attractive for these first applications of Bayesian methods to reaction-theory questions,  more sophisticated methods are needed to describe many reactions of interest. These models may include couplings to collective degrees of freedom, to the continuum, and/or to rearrangement channels. Implementing UQ in these models will require their calibration, and to do that efficiently emulators must be developed. The model-mixing tools discussed in Secs.~\ref{sec:illu_examp} and \ref{subsec:stat_example} are an appealing way to combine treatments of reaction dynamics that are designed for different kinematic domains. BAND's tools will give us the opportunity to leverage these models' local performance in an effort to achieve an overall description of nuclear reactions that is better than that obtained in any individual model.

\section{Case Study: Bayesian Model Averaging in nuclear mass models} 
\label{sec:masses}

The BAND framework will enable quantified  extrapolations
to yet-unexplored domains and to environments that
cannot be directly probed in the laboratory, e.g., the conditions occurring in neutron-star mergers or supernovae. The example below illustrates how the anticipated  BAND tools can enable massive, but still reliable, extrapolations of nuclear properties, such as binding energies.

These extrapolations will establish the limits of nuclear binding and quantify our uncertainty as to where those limits are. This is crucial for understanding how elements in the universe are produced in stellar nucleosynthesis; see, e.g., Ref.~\cite{Horowitz2019}. A quantitative understanding of related astrophysical processes requires knowledge of nuclear properties and reaction rates of thousands of  very exotic  isotopes, the majority of which cannot be accessed by experiments. Consequently, the nuclear data for astrophysical simulations must often be obtained by carrying out massive model-based extrapolations. In several recent studies \cite{Neufcourt2019,Neufcourt2020a,Neufcourt2020b}  BMA techniques were applied to quantify the limits of the nuclear landscape by considering several global models and the most recent experimental information on particle stability  and masses.

The global modeling of all particle-bound nuclei inhabiting the nuclear landscape is a challenging task that  requires  control of many aspects of the nuclear many-body problem. For such a task, the microscopic tool of choice is  nuclear density functional theory based on effective inter-nucleon interactions modeled in terms of energy density functionals (EDFs). Bayesian model calibration has been carried out \cite{McDonnell2015} for some selected EDFs, but not for most of the mass models on the market. In the absence of full uncertainty quantification for each model, a simple and practical strategy \cite{Neufcourt2018, Utama16,Utama18} is to  develop a statistical approach to the residuals  between experimental observations and the predictions of the nuclear mass models across the two-dimensional nuclear domain $\{x_i\}=(Z_i, N_i)$. 
Following the discrepancy approach described in Sec.~\ref{sec:illu_examp}, the Bayesian statistical model for these residuals $y_i - f(x_i,\theta)$ can be written $\delta(x_i)+\varepsilon_{i}$, where  $\delta(x)$ represents the systematic deviation, $\varepsilon$  is the propagated point-to-point uncertainty.
In Refs.~\cite{Neufcourt2019,Neufcourt2020a,Neufcourt2020b}  the function $\delta$ was taken as a GP in the nuclear domain.

\begin{figure}[htb!]
\includegraphics[width=1.0\linewidth]{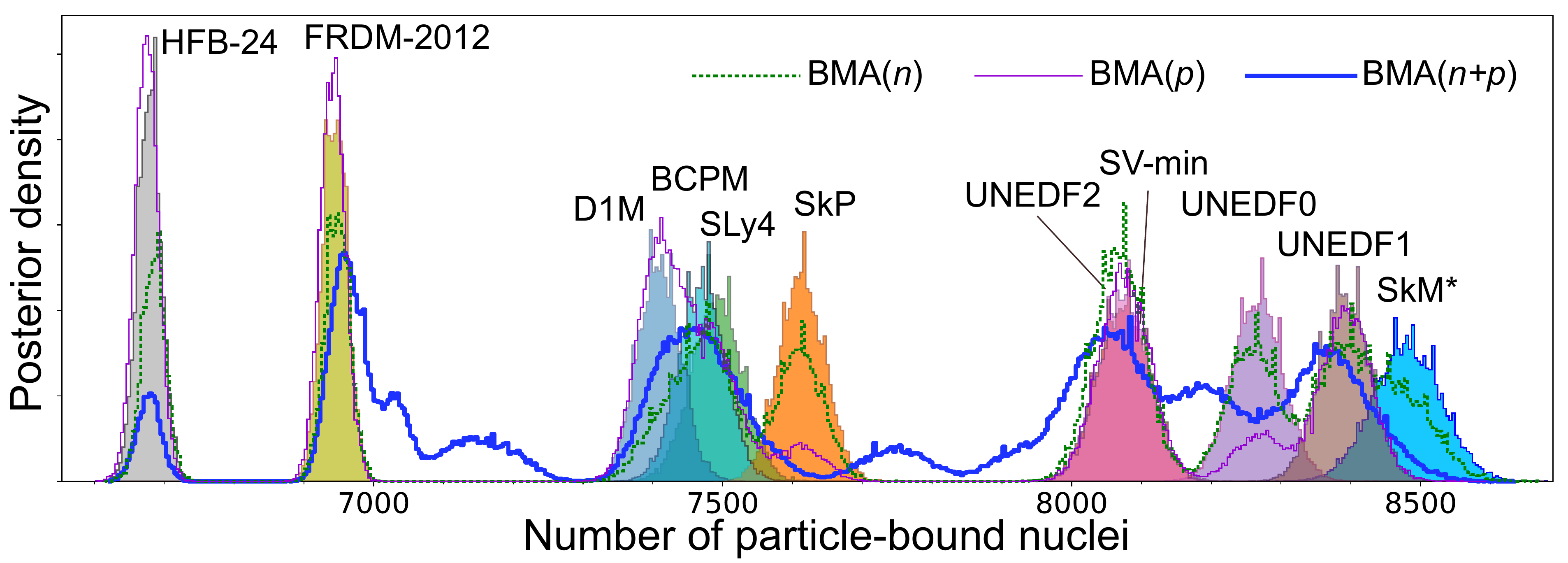}
\caption{
Posterior distributions of the number of particle-bound nuclei with $Z,N\ge 8$ and $Z\le 119$. The histograms show the posterior densities for each model: HFB-24, FRDM-2012, D1M, BCPM, SLy4, SkP, SV-min and UNEDF2, UNEDF0, UNEDF1, and SkM$^*$. The lines show the BMA posterior densities. (From Ref.~\cite{Neufcourt2020b}.)
}
\label{fig:BMA}
\end{figure}

The BMA example presented here is from Ref.~\cite{Neufcourt2020b}, which studied one- and two-nucleon separation energies $S_{1n/1p/2n/2p}$ and particle drip lines. The observations $\observations$ include all experimental masses from atomic mass evaluations AME2003 \cite{AME03b} (training set) together with later measurements from  AME2016 \cite{AME16b} and elsewhere (testing set)~Ref.~\cite{Neufcourt2020b}. The GPs were trained on the separation-energy residuals of $K=11$ nuclear mass models $\mathcal{M}_k$ ($k=1,\dots 11$) that are listed in Fig.~\ref{fig:BMA}. Once the discrepancy functions were inferred in this way, the posterior distributions for each model plus its corresponding discrepancy function were obtained from 50,000 post-burn-in iterations of MCMC. These samples were then used to generate 10,000 mass tables. 
 
The resulting  predictions of the $K=11$ nuclear mass models were then combined via BMA. Ref.~\cite{Neufcourt2020b} used two families of weights based on the data from the neutron-rich ($\design_n $) and proton-rich ($\design_{2p} $)  nuclear domains. On the neutron-rich side, weights were assigned according to the model performance in regard to the prediction of the existence of observed neutron-rich nuclei that were not part of the training or testing sets:
\begin{equation}\label{eq:neutron-weights}
w_k(n)\propto p\left(S_{1n/2n}(x)>0  \text{ for } x\in  \design_n|\mathcal{M}_k \right),
\end{equation}
where $ \design_n$ is the set of 254 experimentally observed neutron-rich nuclei  with $20\le Z \le  50$ for  which no  experimental neutron separation energy is available. On the proton-rich side, weights
\begin{equation}\label{eq:proton-weights}
w_k(p) \propto p\left(S_{2p}(x)<0, S_{1p}(x) > 0 \text{ for } x\in\design_{2p}|\mathcal{M}_k  \right), 
\end{equation}
were given, where $\design_{2p}$ is the set of five long-lived two-proton emitters~\cite{Neufcourt2020a}.
To assess the whole landscape, Ref.~\cite{Neufcourt2020b} applied a local model averaging variant called BMA($n{+}p$),
with local weights that correspond to $w_k(p)$ ($w_k(n)$) on the proton-rich (neutron-rich) side of stability:
\begin{equation}\label{eq:combined-weights}
w_k(Z, N)  = w_k(n)\, H(N{\,\geq\,}N_{\beta}(Z)) + w_k(p)\, H(N{\,<\,}N_{\beta}(Z)),
\end{equation}
with $H(x)$ is the Heaviside step function and $N_{\beta}(Z)$ is 
the neutron number of the average line of $\beta$-stability at proton number $Z$.

To estimate how many particle-bound nuclei  with $Z,N\ge 8$ and  $Z\leq 119$ may exist in nature, the posterior distribution of the number of isotopes with positive separation energies was calculated. The resulting posterior distributions for individual models and BMA are shown in Fig.~\ref{fig:BMA}. According to the BMA($n+p$) analysis in Eq.~(\ref{eq:combined-weights}), the number of particle-bound nuclei is  $7708\pm 534$. The results of the individual models shown in Fig.~\ref{fig:BMA} show considerable spread, primarily due to the extrapolation uncertainty in the heavy neutron-rich region. This result underlines the fact that one should be very careful when trusting extrapolative predictions of any given model.

BAND will take posterior predictions obtained with BMA---such as those discussed in this section---and use them to plan experiments. For this case study those experiments would aim at establishing the existence of exotic nuclei. In the nucleosynthesis context, the  errors on binding energies computed with BMA can guide  the uncertainty analysis for abundance studies involving astrophysical network simulations. BAND will also improve the EDFs used for this study, since full calibration of individual NP models can be considered before they are mixed. Better understanding of the NP model properties in the data space can yield more informed statistical models for the discrepancy between the models and reality than the GP used in the study described above. This, in turn, will permit more robust prediction of extrapolated nuclear properties thus providing better input for experimental design described in Sec.~\ref{sec:design}. 

With BAND, we will improve the simple BMA methodology presented  in this example  by using the more advanced BMM discussed in Sec.~\ref{sec:bmm}. 
In this way, we will be able  to catch local model preferences, see Sec.~\ref{subsec:stat_example} and Ref.~\cite{LeoVojta2020}. Another anticipated improvement concerns the pre-selection of models used in the BMM. This will amount to computing the prior probability $p(\mathcal{M}_k)$ based on the model performance in the space of observations $\design$. This  will enable us to eliminate models that are very similar (or identical) in the space $\design$~\cite{LeoVojta2020}. 

\section{Case Study: Bayesian Model Averaging for transport coefficients in dynamical models of heavy-ion collisions}
\label{sec:BMA_TC}

A simple application of Bayesian Model Averaging to heavy-ion collisions dynamics was recently published by the JETSCAPE Collaboration \cite{Everett:2020yty}. One of JETSCAPE's goals is to use experimental data measured at RHIC and the LHC to perform global calibration of a highly complex dynamical model for the evolution of hot and dense quantumchromodynamics (QCD) matter created in relativistic heavy-ion collisions \cite{Everett:2020xug}. There is, however, an irreducible model uncertainty in the calibration. It arises from ambiguities in the model used for ``particlization". Particlization marks the transition between two dynamical modules: a relativistic dissipative fluid dynamical description of the early quark-gluon plasma stage of the heavy-ion collision and a microscopic kinetic transport code describing the late and much more dilute hadronic stage. Particlization is necessary to translate the fluid from the first stage into the set of particles that get transported in the second stage. The posterior joint probability distribution $\mathcal{P}(\theta|\mathbf{y}_{\exp})$ for 17 model
%
\begin{figure}[h]
\centering
\includegraphics[width=0.8\linewidth]{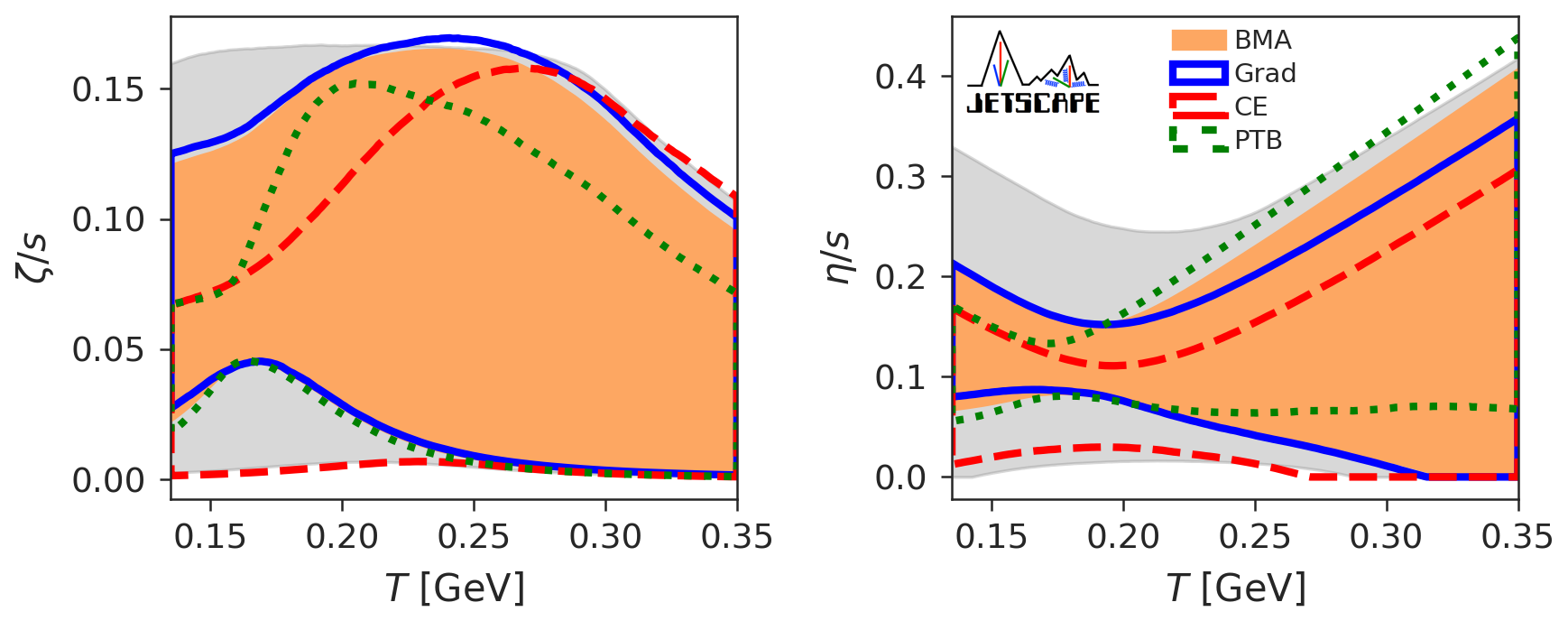}
\caption{The $90$\% credible intervals for the prior (gray), the posteriors of the Grad (blue), Chapman-Enskog (red) and Pratt-Torrieri-Bernhard (green) particlization models, and their Bayesian model average (orange) for the specific bulk (left) and shear (right) viscosities of QGP. (From Ref.~\cite{Everett:2020yty}.)
}
\label{fig:posterior_credible_interval}
\end{figure}
%
parameters $\theta$ was extracted via Bayesian Model Averaging. As in Eq.~(\ref{eqn:posteriorBMA}),  that posterior is a linear combination of the posteriors corresponding to three different model choices for this transition. These models are denoted Grad, PTB (Pratt-Torrieri-Bernhard), and CE (Chapman-Enskog) in Fig.~\ref{fig:posterior_credible_interval}.
For the case studied in \cite{Everett:2020yty} the evidence ratios of these models were approximately 5000:3000:1; that is, the CE model turned out to be significantly disfavored by the data while the other two contributed with similar weights to the Bayesian Model Average. The resulting 90\% credibility intervals for the specific shear and bulk viscosities, $\eta/s$ and $\zeta/s$, as functions of temperature are shown in Fig.~\ref{fig:posterior_credible_interval}. The gray areas denote the prior 90\% credible intervals (see Ref.~\cite{Everett:2020xug} for an in-depth discussion of prior selection), the colored lines outline the corresponding ranges for the three particlization models studied in \cite{Everett:2020yty, Everett:2020xug}, while the orange areas show the ones for the Bayesian Model Averages. The differences between the prior (gray) and posterior (orange) 90\% credible intervals for the Quark-Gluon Plasma (QGP) viscosities indicate that the available experimental data exhibit their strongest constraining power in the lower temperature region 150\,MeV${\ \lesssim\ }T{\ \lesssim\ }250$\,MeV; above $T{\,\approx\,}250$\,MeV their power to constrain these transport coefficients rapidly degrades, leaving large uncertainties for both the shear and, in particular, the bulk viscosity. For a deeper discussion of the physical and statistical implications of this plot we refer the reader to \cite{Everett:2020yty}. 

The study presented in Refs.~\cite{Everett:2020yty, Everett:2020xug} employed a number of tools used in Bayesian inference that are anticipated to become, in one form or another, part of the BAND framework. This will facilitate their application to a much wider set of problems in Nuclear Physics: (i) economic sampling of a high-dimensional model parameter space using a Latin hypercube design for full model runs; (ii) Principal Component Analysis (PCA) of a large space of observables to reduce the dimensionality of the space of target observables for calculating the likelihood of the model parameters; (iii) GP emulators  trained on the PCA observables predicted by the full-model runs to efficiently interpolate these predictions to large numbers of alternate model parameter settings; (iv) closure tests for testing emulator performance and our ability to reconstruct the model parameters from ``mock data" generated by the full model with known parameter settings; (v) efficient MCMC sampling of the multidimensional posterior probability distribution for the model parameters; and (vi) Bayesian Model Averaging  to combine the posterior distributions from different, a priori equally likely models, in order to quantify the contribution of irreducible model uncertainties to the variance of parameters inferred from the experimental data. In developing these tools and applying them appropriately, collaboration between physicists and statisticians has been invaluable, and BAND will follow the same strategy.

A key deliverable of the BAND initiative is a statistically meaningful simultaneous quantification of both theoretical and experimental uncertainties in Bayesian inference. The study reported in Refs.~\cite{Everett:2020yty, Everett:2020xug} made a first step in this direction within the context of heavy-ion collision dynamics. But its scope was limited because it considered only the theoretical uncertainty associated with the particlization of the quark-gluon plasma fluid {\it at the end of its evolution}. As mentioned in Sec.~\ref{sec:eos}, other modeling uncertainties affect {\it the early evolution stages} and even the initial conditions of QCD matter created in heavy-ion collisions. For studying the interplay of early and late modeling uncertainties, and the best weighting of these in future predictions of additional observables for experimental design, the discussion presented in Sec.~\ref{sec:bmm} clarifies that the simple linear combination of the posterior distributions of each individual model used in Ref.~\cite{Everett:2020yty} is no longer adequate. The BAND initiative will combine expertise in physics, statistics and computer science to develop and implement more powerful Bayesian Model Mixing tools needed to properly account for local model preferences while also adequately accounting for the individual models' overall performance in the space of observations $\observations$ through their model evidence $p(\mathcal{M}_k)$.

\section{Strike up the BAND} \label{sec:conclusion}

The BAND framework is designed to be an integrated set of computational and input tools. The BAND collaboration will develop the framework in several stages that will include concurrent lines of development and testing. Open-source code development and delivery will be facilitated via the BAND Github repository~\cite{BANDGithub}. We will develop codes for novel applications using a mix of the repository's public and private branches. The framework will also draw on and integrate other repositories where publicly available open-source codes that perform BAND-relevant physics and statistics functions reside. The BAND framework will be intentionally permissive in terms of the languages and formats of collaboration code. The computational/theoretical models that can be interfaced with BAND framework codes will thus range in language (e.g., Fortran, C/C++, Python) and scale (e.g., executable on a single thread, with its own MPI communicator). This fusion of disparate tools will be achieved by adhering to newly designed BAND Software Development Kit (SDK) requirements. This SDK will borrow from established community software requirements such as those of the Extreme-scale Scientific Software Development Kit (xSDK) \cite{xSDK} and IDEAS Productivity \cite{IDEAS} efforts.  The goal of this SDK is to build in interoperability across the BAND software ecosystem, large-scale scientific simulation codes, and other numerical libraries.  This will enable non-BAND scientists' involvement in the development of BAND's instruments and in proof-of-concept science analyses. 

BAND is already collating, documenting, and linking to or storing codes from various sub-fields of nuclear physics. New framework codes will be developed in parallel with interfaces that allow the use of existing modeling code within the framework. For example, the model calibration component of BAND will involve new technology for emulation and posterior exploration that interfaces with existing  GP emulators and MCMC methods. The resulting capabilities will be part of the first release of the framework, scheduled for 2021. That release will have limited physics functionality but serve as a testing platform. Unit and regression tests will be used to ensure that core functionalities are maintained during BAND's continuous, community-oriented development. Later releases will include the entire suite of tools depicted in Fig.~\ref{fig:flowchart}. All releases will be available for download from our public repository, so any interested community member can test and develop familiarity with the evolving framework.  

Nuclear physicists will then be able to bring their physics model and dataset and use BAND's input tools to:
\begin{itemize}
    \item Formulate a likelihood. Section~\ref{subsec:likelihood} explains the Bayesian approach to formulating likelihoods that users can employ for parameter estimation and making predictions. BAND will encourage them to consider error modeling that goes beyond the standard likelihood (\ref{eq:standardL}) in order to account for deficiencies in their physics model.
    
    \item Specify priors. BAND's participatory approach to prior selection, discussed in Sec.~\ref{subsec:priors}, will facilitate the development of priors that encode physical bounds on parameters, or expectations regarding their natural size. This will mean that {\it all} pertinent information, not just that in the provided dataset, will be leveraged and accounted for in the posteriors for {\it all} quantities of interest.
\end{itemize}
Of course, the statistical models developed in this way must be checked. BAND will employ a number of statistical model-checking diagnostics (see, e.g., Ref.~\cite{BastosOHagan} for the GP case) to ensure that the statistical models adopted are consistent. We will particularly focus on whether the BAND framework produces accurate credibility intervals, i.e., the 68\% credibility interval around the model prediction encompasses the correct result 68\% of the time.  
    
BAND's inter-operable computational tools will also facilitate model emulation, which is crucial for NP models that require large amounts of computer time for a single evaluation. BAND's emulators will then be used to map out the posterior via Monte Carlo sampling. In this way, BAND can be used for efficient calibration of a single NP model. 

But a key emphasis of BAND is to go beyond such a single-model approach and use Bayesian Model Mixing to obtain more information---and more reliable information---than is available in the posterior of any one NP model. The principles of BMM were explained in Sec.~\ref{sec:bmm}. BMM can be superior to Bayesian Model Averaging because it does not generate the full posterior of each model before averaging them, but instead employs more specific information on each model to produce a posterior that draws on each model in its areas of strength. 

Section~\ref{sec:illu_examp} applied the emulation, calibration, and Bayesian Model Mixing elements of the BAND framework in a simple context: the problem of estimating the gravitational acceleration from data in a ball-drop experiment. 

The results of BAND analyses---whether single- or multi-model---will then be used to perform experimental design analyses, i.e., answer questions about what experiment will produce the maximum gain in regard to a desired piece (or pieces) of information---see Sec.~\ref{sec:design}. 

Finally, in Secs.~\ref{sec:eos}--\ref{sec:BMA_TC} we discussed some recent applications of Bayesian methods in NP and explained how the BAND framework will enable analyses that go much further. BAND's ability to develop statistical models of the discrepancy between physics models and data, together with its intelligent use of priors, and its emphasis on Bayesian Model Mixing, will provide deeper insights into the equation of state, initial conditions and transport coefficients of strongly interacting matter, the existence of nuclei near the driplines, production of elements in stars, and models of nuclear reactions. In each area BAND's full quantification of uncertainties will allow it to provide valuable guidance regarding the impact of proposed experiments at FRIB, RHIC, and other NP facilities.
 
\section*{Acknowledgments}
We thank Derek Everett and Pablo Guiliani for reading the manuscript carefully and suggesting several improvements.
This work was supported by the National Science Foundation CSSI program under award number OAC-2004601 (BAND Collaboration). Additional support was provided in part by the National Science Foundation under award numbers NSF PHY-1913069 (R.J.F.), ACI-1550223 (U.H., within the framework of the JETSCAPE Collaboration), NSF-DMS-1953111 (M.P.), NSF-PHY-1811815 (F.N.), and in part by the U.S. Department of Energy, Office of Science, Office of Nuclear Physics under award numbers 
\rm{DE-SC0013365} (W.N.), \rm {DE-FG02-03ER41259} (S.P.), \rm{DE-SC0004286} (U.H.), \rm{DE-FG02-93ER40756} (D.R.P.), 
Office of Advanced Scientific Computing Research under contract number DE-AC02-06CH11357 (S.M.W.), and the NUCLEI SciDAC project. 
U.H. acknowledges support by the Alexander von Humboldt Foundation through a Humboldt Research Award. 

\section*{References}
\providecommand{\newblock}{}

\end{document}